\documentclass[twocolumn,amsmath, trackchanges]{aastex702}
\usepackage{etoolbox}

\makeatletter
\journalinfo{}
\patchcmd{\frontmatter@title@above}
{{\footnotesize Typeset using \LaTeX\ {\bf twocolumn} style in AASTeX7.0.2}}
{}
{}
{\PackageWarning{aastex702}{Failed to remove Typeset information}}
\makeatother

\usepackage{graphicx,latexsym,amssymb,epsfig}
\usepackage{multirow,amsmath,array,booktabs}
\usepackage{natbib}
\usepackage{url}
\usepackage{subfigure}
\usepackage{tabularx}
\usepackage{color}
\usepackage{bm}
\usepackage{color}
\usepackage{braket}
\usepackage[figuresright]{rotating}

\makeatletter
\renewcommand{\@keys@name}{%
	\textit{Unified Astronomy Thesaurus concepts:}\/~\mbox{}%
}
\makeatother
\begin{document}

\title{Impact of effective nucleon mass and symmetry energy on nuclear pasta phases and the unified equation of state of neutron stars}

\author[gname=Shuying, sname=Li]{Shuying Li}
\affiliation{School of Physics, Nankai University, Tianjin 300071, People’s Republic of China}
\email{li_shu_ying_ll@163.com}

\author[0000-0003-2717-9939, gname=Hong, sname=Shen]{Hong Shen}
\affiliation{School of Physics, Nankai University, Tianjin 300071, People’s Republic of China}
\email{shennankai@gmail.com}

\author[0000-0002-1709-0159, gname=Jinniu, sname=Hu]{Jinniu Hu}
\affiliation{School of Physics, Nankai University, Tianjin 300071, People’s Republic of China}
\email{hujinniu@nankai.edu.cn}

\begin{abstract}
We investigate how the effective nucleon mass and symmetry energy affect pasta phase properties 
and the unified equation of state (EOS) of neutron stars.
The nonuniform pasta structure is described within the self-consistent Thomas--Fermi approximation,
while nucleonic interactions are treated in the relativistic mean-field framework
with several typical parameter sets adopted. 
Our results indicate that the symmetry energy and its slope are crucial in determining the onset
of pasta phases and their geometric configurations. In the high-density region where matter 
becomes uniform, the effective nucleon mass plays a dominant role. 
A larger effective nucleon mass tends to soften the EOS, which yields a smaller radius 
for a canonical 1.4 $M_{\odot}$ neutron star and reduces the corresponding tidal deformability.
\end{abstract}
\keywords{\uat{Neutron stars}{1108};\uat{Compact objects}{288};
	\uat{Nuclear astrophysics}{1129}	
}
\section{Introduction} \label{sec:1}

Neutron stars are among the most extreme forms of matter in the universe, with 
interior densities spanning from subnuclear values to several times the nuclear 
saturation density. Consequently, they are regarded as a unique natural 
laboratory for studying the properties of dense matter~\citep{latt04}.
With advances in astronomical observation techniques, it is now possible to test 
theoretical models of nuclear matter using multiple observational constraints from 
neutron stars, such as the mass-radius relation, tidal deformability, 
and rotation rate~\citep{latt07, latt14a, marg21}.
In the study of neutron stars, the equation of state (EOS) plays a crucial role 
in determining the observable properties of neutron stars.
The neutron-star EOS covers a wide density range, which is generally divided into three regions: 
the outer crust, the inner crust, and the liquid core.
The density in the outer crust is relatively low, falling below the neutron drip density 
of about $2\times 10^{-4} \, \text{fm}^{-3}$~\citep{baym71b,bao14b,oert17}. 
It is composed of spherical nuclei embedded in a background electron gas~\citep{rust06}.
The inner crust spans from the neutron drip density to the crust-core transition 
at about $1/2$ nuclear saturation density $n_0$.
As the density increases, nuclei in the inner crust become increasingly neutron-rich, coexisting 
with free neutrons and electrons. 
Furthermore, spherical nuclei may become unstable and transform into nonspherical geometries, 
including rods, slabs, tubes, and bubbles~\citep{rave83,oyam07}. 
These configurations are known as nuclear pasta phases, arising from the competition 
between nuclear surface and Coulomb energies~\citep{bao14a,bao15,scur23}.
Owing to the complexity of pasta phases, constructing the inner crust EOS is a challenging task, 
with outcomes being highly sensitive to the approximate method and nuclear model employed.
Beyond the crust-core transition density, the stable phase is homogeneous matter in $\beta$ 
equilibrium, consisting of neutrons, protons, and leptons.
The core of a neutron star is composed of homogeneous matter and dominates its mass and radius.
Non-nucleonic degrees of freedom in the dense interior of neutron stars have been widely 
discussed in the literature~\citep{webe05,oert17}.
In general, these exotic components lead to a significant reduction 
in the maximum neutron-star mass~\citep{ju21,huang22b,yama22}.
For simplicity, we do not include non-nucleonic degrees of freedom in the present study. 

The primary challenge in exploring the inner crust is to properly describe the complex phase structure 
of nonuniform matter. There are several methods to address this challenge, such as 
the coexisting phases (CP) method~\citep{bao14b,wang22}, 
the compressible liquid-drop (CLD) model~\citep{bao14b,carr19,scur23}, 
and the Thomas--Fermi (TF)  approximation~\citep{goge07,avan09,avan10,okam13,ji21,xia21}.
Among these methods, the TF approximation can provide a relatively realistic description of
nuclear pasta phases, where the density distributions are determined self-consistently.
In \citet{goge07}, a parameterized TF approximation was employed to describe 
inhomogeneous nuclear matter in neutron-star crusts, using both the Skyrme model and 
the relativistic mean-field (RMF) approach for nuclear interactions. 
A detailed comparison was made between the results of Skyrme–Hartree–Fock calculations and RMF calculations. 
In \citet{avan09,avan10}, a self-consistent TF approximation was adopted 
to study the properties of pasta phases at both finite and zero temperatures,
and was compared with the results obtained from the CP, spinodal, and binodal methods. 
It has been shown that the crust-core transition density is sensitive to the method used.
The dynamical spinodal defines a lower limit for this transition density while the binodal 
defines a higher limit, and the TF result lies between these limits. 
In addition, the TF approximation has been applied to perform three‑dimensional calculations
for pasta structures~\citep{okam13,ji21,xia21}, and the influence of symmetry energy 
has also been investigated.
Going beyond the semiclassical TF framework, three-dimensional Skyrme-Hartree-Fock calculations 
without imposing specific pasta geometries have been performed using 
several Skyrme parametrizations~\citep{newt09,pais14,fatt17}. 
Various intermediate configurations connecting the canonical pasta phases were found, 
which could help smooth the transitions between different pasta shapes.

Recently, considerable attention has been paid to the impact of effective nucleon mass $M^{\ast}$ 
on supernova explosions in the astrophysics community~\citep{schn19,yasi20}.
The value of $M^{\ast}$ varies significantly across different theoretical frameworks.
In nonrelativistic Skyrme models, the effective mass ratio $ M^{\ast}/M $ at saturation density 
is often treated as a model parameter, ranging from $0.5$ to $1.0$.
On the other hand, in RMF models, the effective nucleon mass is given by
$ M^{\ast} = M + g_\sigma \sigma$, which is lower than the free nucleon mass $M$ 
due to the large negative scalar potential $U_S=g_{\sigma}\sigma$.
The saturation mechanism of symmetric nuclear matter in RMF models arises from
a delicate balance between the large negative scalar and positive vector potentials.
Accordingly, the ratio $ M^{\ast}/M $ at saturation density in RMF models generally
falls within a narrow range of approximately $0.54-0.8$.
It has been reported in~\citet{schn19,yasi20} that the effective nucleon mass
plays a decisive role in supernova explosions, affecting the pressure,
proto-neutron star contraction, and neutrino emission.
Specifically, a larger effective mass leads to smaller thermal contributions to the pressure,
which results in more rapid contraction and accelerates shock evolution toward a faster explosion.
In their calculations, the influence of the effective mass $M^{\ast}$ was investigated using 
a set of Skyrme-type EOSs by varying its value at saturation density, 
which was treated as a model parameter within a nonrelativistic framework.
Moreover, the influence of $M^{\ast}$ on binary neutron-star mergers has been explored within 
a nonrelativistic framework through numerical simulations~\citep{fiel23,rait23}.
It has been shown in~\citet{fiel23} that a larger $M^{\ast}$ enhances the specific 
heat capacity, thereby producing a compact merger remnant with lower temperatures 
and shifting the peak frequency of the postmerger gravitational-wave signal to higher values.
In our previous work~\citet{li25a,li25b}, we extended the study of the effective nucleon mass 
within a relativistic framework. Based on the RMF approach with TM1e parametrization~\citep{bao14b,shen20}, 
we proposed two new parameter sets, TM1m* and TM1m, both retaining saturation properties similar
to those of TM1e but with a larger effective nucleon mass $M^{\ast}/M \sim 0.8$. 
In~\citet{li25a}, we investigated the influence of the effective nucleon mass 
on the EOS at higher densities for supernova simulations. In addition, a comparison between 
the relativistic TM1m model and the nonrelativistic Skyrme model was presented.
Furthermore, we proposed the TM1m* model to improve the description of finite nuclei, 
while retaining saturation properties similar to those of the TM1m model~\citep{li25b}.
In principle, the effective nucleon mass should be density-dependent and can be incorporated 
into a relativistic framework more consistently than in nonrelativistic approaches.

This article aims to extend our investigation of the effective nucleon mass to pasta phases 
occurring in the inner crust of neutron stars. In addition, we construct several unified EOSs by
employing the same nuclear model for both the inner crust and the core of neutron stars, 
to examine the influence of $ M^{\ast} $ on neutron-star properties.
We use the TF approximation to describe inhomogeneous nuclear matter,
as its energy functional naturally incorporates all necessary components and 
self-consistently yields a realistic description of nuclear pasta phases.
To examine the impact of the effective mass and symmetry energy, we adopt six RMF models:
TM1m*~\citep{li25b}, TM1m~\citep{li25a}, TM1e~\citep{shen20}, TM1~\citep{suga94}, 
NL3~\citep{lala97}, and SFHo~\citep{stei13}. 
Among these models, the TM1m*, TM1m, and TM1e models are all derived from the original TM1,
collectively referred to as the TM1-family models. 
The TM1m* and TM1m models have a larger effective mass of $M^{\ast}/M \sim 0.8$, 
while the TM1e and TM1 models have relatively smaller values of $M^{\ast}/M \sim 0.63$.
The difference between TM1e and TM1 lies in the density dependence of symmetry energy, 
reflected in the symmetry energy slopes of $L=40$ MeV (TM1e) and $L=111$ MeV (TM1).
In the TM1-family models, the isoscalar saturation properties, 
such as the saturation density $n_0$, energy per particle $E/A$, and incompressibility $K$,
remain the same, while the effective mass $M^{\ast}/M$ and behaviors of symmetry
energy ($E_{\mathrm{sym}}$ and $L$) are markedly different.
By comparing the results of TM1m* and TM1m with those of TM1e, we can analyze the impact 
of the effective mass, whereas the comparison between TM1e and TM1 can reveal the influence 
of symmetry energy $E_{\mathrm{sym}}$ and its slope $L$.
It is well known that the symmetry energy slope $L$ can significantly affect neutron-star 
properties, such as the mass-radius relation and tidal deformability~\citep{fatt13,latt14b}. 
The pasta structure and crust-core transition are also influenced by the symmetry energy 
slope $L$~\citep{bao15,ji19}. 
For further examination of model dependence, we include two other widely used models: 
the NL3 model with a large slope parameter of $L=118$ MeV, comparable to TM1, 
and the SFHo model known for its relatively large effective mass of $M^{\ast}/M \sim 0.76$ 
and wide applications in astrophysical research.

This article is organized as follows. In Section~\ref{sec:2}, we briefly describe the RMF models
employed in this study and introduce the self-consistent TF approximation to describe 
pasta phases in the inner crust of neutron stars. In Section~\ref{sec:3}, we present numerical 
results for pasta phases and analyze the impact of effective nucleon mass and symmetry energy 
on their properties. In addition, we investigate neutron-star properties using unified EOSs.
Finally, in Section~\ref{sec:4}, we provide a summary and present our conclusions.

\section{Model and Parameters} \label{sec:2}

We employ the RMF model to construct the unified EOS of neutron stars, 
including the uniform core and the nonuniform inner crust.
In the RMF approach, the interaction between nucleons is described in terms of the exchange of 
the isoscalar-scalar meson $\sigma$, the isoscalar-vector meson $\omega$, and 
the isovector-vector meson $\rho$.
For cold neutron-star matter, we include neutrons, protons, and electrons, as well as muons 
that may appear at higher densities. 
The Lagrangian density of this system reads
\begin{equation}
	\label{Lagrangian}
	\begin{aligned}
		\mathcal{L}_{\mathrm{RMF}}
		=& \sum_{i=p, n} \bar{\psi}_i\Big\{i \gamma_\mu  \partial^\mu-\left(M+g_\sigma \sigma\right) \\
		& \left.-\gamma_\mu\left[g_\omega \omega^\mu+\frac{g_\rho}{2} \tau_a \rho^{a \mu}+\frac{e}{2}\left(1+\tau_3\right) A^\mu\right]\right\} \psi_i	\\
		& +\frac{1}{2} \partial_\mu \sigma \partial^\mu \sigma-\frac{1}{2} m_\sigma^2 \sigma^2-\frac{1}{3} g_2 \sigma^3-\frac{1}{4} g_3 \sigma^4 \\
		& -\frac{1}{4} W_{\mu \nu} W^{\mu \nu}+\frac{1}{2} m_\omega^2 \omega_\mu \omega^\mu+\frac{1}{4} c_3\left(\omega_\mu \omega^\mu\right)^2 \\
		& -\frac{1}{4} R_{\mu \nu}^a R^{a \mu \nu}+\frac{1}{2} m_\rho^2 \rho_\mu^a \rho^{a \mu} \\
		& +\Lambda_{\mathrm{v}}\left(g_\omega^2 \omega_\mu \omega^\mu\right)\left(g_\rho^2 \rho_\mu^a \rho^{a \mu}\right)-\frac{1}{4} F_{\mu \nu} F^{\mu \nu} \\
		& +\sum_{j=e, \mu} \bar{\psi}_j\left[i \gamma_\mu \partial^\mu-m_j+e \gamma_\mu A^\mu\right] \psi_j ,
	\end{aligned}
\end{equation}
where $W^{\mu\nu}$, $R^{a\mu\nu}$ and $F^{\mu\nu}$ denote the antisymmetric field tensors corresponding 
to $\omega^{\mu}$, $\rho^{a\mu}$ and $A^{\mu}$, respectively. 
Within the mean-field approximation, these meson fields are treated as classical fields 
by replacing the operators with their non-zero expectation values:
$\sigma=\braket{\sigma}$, $\omega=\braket{\omega^{0}}$, and $\rho=\braket{\rho^{30}}$. 
The chemical potentials of nucleons are given by
\begin{align}
	\mu_p &= \sqrt{\left(k_F^p\right)^2+M^{\ast 2}}+g_\omega \omega+\frac{g_\rho}{2} \rho+e A,
	\label{chem_pot_p} \\
	\mu_n &= \sqrt{\left(k_F^n\right)^2+M^{\ast 2}}+g_\omega \omega-\frac{g_\rho}{2} \rho,
	\label{chem_pot_n}
\end{align}
where $k^{i}_{F}$ is the Fermi momentum of species $i$, and the corresponding number density 
is given by $n_{i}=(k^{i}_{F})^{3}/3\pi^2$. 
Here, $M^{\ast}=M+g_{\sigma}\sigma$ denotes the effective nucleon mass, which is determined 
by the negative scalar potential $U_S=g_{\sigma}\sigma$.

The neutron-star core consists of homogeneous $npe\mu$ matter under the conditions of $\beta$ 
equilibrium and charge neutrality. The core EOS is obtained by solving a set of coupled 
equations self-consistently within the RMF approach~\citep{bao15,ji19,li25a}.
As the density decreases, uniform matter becomes energetically unfavorable 
compared to cluster formation, and the system undergoes a crust-core transition.
Correspondingly, nucleons cluster into pasta phases or spherical nuclei in the inner 
crust of neutron stars. The unified EOS employs the same RMF model for both the core and the inner crust.

To describe nonuniform matter in the inner crust, we perform a self-consistent TF calculation 
as described in our previous work~\citet{bao15}. The Wigner--Seitz cell approximation is 
adopted to simplify the treatment of pasta phases. Inside a unit cell, neutrons and protons 
display inhomogeneous distributions, obtained self-consistently within the TF approximation.
For simplicity, the electron density is assumed to be uniform throughout the cell,
since the electron screening effect is negligible at subnuclear densities~\citep{maru05}.
The volume of the Wigner--Seitz cell with different configurations is expressed as
\begin{eqnarray}
	V_{\text {cell }}= \begin{cases}\frac{4}{3} \pi r_{\mathrm{ws}}^3 & (\text{droplet and bubble}), \\ l \pi r_{\mathrm{ws}}^2 & (\text {rod and tube}), \\ 2 r_{\mathrm{ws}} l^2 & (\text{slab}),\end{cases}
\end{eqnarray}
where $r_{\mathrm{ws}}$ denotes the radius of a spherical cell in the droplet and bubble configurations. 
In the rod and tube configurations, the cell is cylindrical with radius $r_{\mathrm{ws}}$ and length $l$, 
while in the slab configuration, the cell has width $l$ and thickness $2 r_{\mathrm{ws}}$. 
It is noteworthy that the choices of cell length for cylindrical cells and width for slabs are somewhat
arbitrary, and such choices do not affect the resulting energy density of the system~\citep{gril12}. 

In the TF approximation, the nucleon density distribution varies continuously throughout the cell,
without a sharp interface between the nucleus and the surrounding neutron gas, in contrast to 
the liquid-drop model. We therefore employ the method described in \citet{maru05} 
to calculate the size of the inner part by density fluctuations,
\begin{equation}
	\label{rd}
	r_d= \begin{cases}r_{\mathrm{ws}}\left(\frac{\langle n_p \rangle ^2}{\left\langle n_p^2\right\rangle}\right)^{1 / D} & (\text {droplet, rod, and slab}), \\ r_{\mathrm{ws}}\left(1-\frac{\langle n_p\rangle ^2}{\left\langle n_p^2\right\rangle}\right)^{1 / D} & (\text {tube and bubble}),\end{cases}
\end{equation}
where $D=1,2,3$ denotes the geometrical dimension of the pasta structure.

At a given baryon density $n_{b}$, the spatial distributions of protons and neutrons in each cell,
described by $n_p(\mathbf{r})$ and $n_n(\mathbf{r})$, satisfy the constraint of baryon number conservation,
\begin{equation}
	\label{numcon}
	\int_{\mathrm{cell}}\left[n_p(\mathbf{r})+n_n(\mathbf{r})\right] 
	d \mathbf{r} \bigg/  V_{\mathrm{cell}} = n_b.
\end{equation}
In addition, the constraints of $\beta$ equilibrium and charge neutrality are imposed within
the cell,
\begin{align}
	\mu_n &= \mu_p + \mu_e,  
	\label{betaeq} \\
	\int_{\text{cell}} n_p(\mathbf{r}) \, d\mathbf{r} & = n_e V_{\mathrm{cell}}. 
	\label{charge}
\end{align}
With these constraints, we solve the meson field equations self-consistently to obtain the field 
profiles and nucleon density distributions. Subsequently, the total energy per cell is calculated by
\begin{equation}
	\label{ecell}
	E_{\mathrm{cell}}=  \int_{\mathrm{cell}} \varepsilon_{\mathrm{RMF}}(\mathbf{r}) d \mathbf{r}+\varepsilon_e V_{\mathrm{cell}},
\end{equation}
where $\varepsilon_{e}$ is the kinetic energy density of electrons. 
Here, $\varepsilon_{\mathrm{RMF}}(\mathbf{r})$ denotes the local energy density in the RMF model, given by
\begin{equation}
	\begin{aligned}
		\varepsilon_{\mathrm{RMF}}(\mathbf{r})= & \sum_{i=p, n} \frac{1}{\pi^2} \int_0^{k_F^i} d k  \, k^2 \sqrt{k^2+M^{\ast 2}} \\
		& +\frac{1}{2}(\nabla \sigma)^2+\frac{1}{2} m_\sigma^2 \sigma^2+\frac{1}{3} g_2 \sigma^3+\frac{1}{4} g_3 \sigma^4 \\
		& -\frac{1}{2}(\nabla \omega)^2-\frac{1}{2} m_\omega^2 \omega^2-\frac{1}{4} c_3 \omega^4+g_\omega \omega\left(n_p+n_n\right) \\
		& -\frac{1}{2}(\nabla \rho)^2-\frac{1}{2} m_\rho^2 \rho^2-\Lambda_{\mathrm{v}} g_\omega^2 g_\rho^2 \omega^2 \rho^2 \\
		& +\frac{g_\rho}{2} \rho\left(n_p-n_n\right)-\frac{1}{2}(\nabla A)^2+e A\left(n_p-n_e\right).
	\end{aligned}
\end{equation}

In practice, at a given average baryon density $n_b$, we start with a proper initial guess 
for the mean fields $\sigma(\mathbf{r})$, $\omega(\mathbf{r})$, $\rho(\mathbf{r})$, and $A(\mathbf{r})$,
then determine the chemical potentials $\mu_n$, $\mu_p$, and $\mu_e$ under the 
constraints of Equations.~(\ref{numcon})-(\ref{charge}). Using these chemical potentials,
we calculate various densities and solve the meson field equations again to obtain new mean
fields. This procedure is iterated until convergence is achieved.
For a given pasta configuration, the cell size $r_{\mathrm{ws}}$ is determined by minimizing 
the energy density $\varepsilon = E_{\mathrm{cell}} / V_{\mathrm{cell}}$.
These TF calculations are carried out for all five configurations considered. 
We compare their energy densities and select the configuration with the lowest one as the 
most favorable pasta configuration at the density $n_b$. 
When the energy density of pasta phases exceeds that of uniform matter,
the system undergoes the crust-core transition. According to thermodynamic relations,
the pressure $P$ of pasta phases is derived from the energy density $\varepsilon$ as
\begin{eqnarray}
	P=\sum_{i=n,p,e} \mu_i n_i-\varepsilon.
\end{eqnarray}

For the nuclear interaction, we consider six RMF models:
TM1m*, TM1m, TM1e, TM1, NL3, and SFHo, in order to examine the impact of the effective nucleon 
mass and symmetry energy. The saturation properties of these six models are listed in Table~\ref{properties}. 
The TM1m* and TM1m models are derived from TM1e by raising the effective nucleon mass ratio $M^{\ast}/M$ 
at saturation density from 0.63 to approximately 0.8, while keeping the other saturation properties 
unchanged. Using these two models, we can investigate the influence of effective nucleon mass.
The main difference between TM1m* and TM1m lies in their description of finite nuclei. 
Specifically, the TM1m* model incorporates additional experimental constraints from binding energies 
and charge radii of several finite nuclei during the parameter fitting. As a result, the TM1m* model
provides an improved description of finite nuclear properties~\citep{li25b}.
The TM1e model is an extension of the original TM1 model, obtained by modifying the isovector coupling 
parameters $g_{\rho}$ and $\Lambda_{\mathrm{v}}$, thereby reducing the symmetry energy $E_{\mathrm{sym}}$ 
and its slope $L$ at saturation density. 
For convenience, we list the parameter sets for the TM1-family models in Table~\ref{parameter}.
It can be seen that TM1m* and TM1m have relatively smaller $g_{\sigma}$ and $g_{\omega}$, 
which result in larger effective masses than those of TM1e and TM1.
The NL3 model was determined by fitting experimental data of finite nuclei.
It is seen in Table~\ref{properties} that the symmetry energy slopes of TM1 and NL3
are much larger than those of the other models. The SFHo model is widely used in astrophysical research, 
and its effective mass ratio $M^{\ast}/M=0.76$ is close to those of the TM1m* and TM1m models. 

\begin{table*}[tbp]
\begin{center}
	\caption{Saturation properties of nuclear matter in the TM1-family (TM1m*, TM1m, TM1e and TM1), NL3, and SFHo models. The table presents the saturation density $n_0$, energy per particle $E/A$, incompressibility $K$, symmetry energy $E_{\mathrm{sym}}$ and its slope $L$, and effective nucleon 
		mass ratio $M^*/M$.}
	\label{properties}
	\begin{tabular}{>{\centering\arraybackslash}p{3cm} >{\centering\arraybackslash}p{2cm} >{\centering\arraybackslash}p{2cm} >{\centering\arraybackslash}p{2cm}
	>{\centering\arraybackslash}p{2cm} >{\centering\arraybackslash}p{2cm} 
    >{\centering\arraybackslash}p{2cm}}
		\hline\hline
		& \textbf{TM1m*} & \textbf{TM1m} & \textbf{TM1e} & \textbf{TM1} & \textbf{NL3} &\textbf{SFHo}   \\
		\hline
		$n_0\,(\text{fm}^{-3})$ & 0.145&0.145&0.145&0.145 & 0.148 &0.158\\
		$E/A\,\text{(MeV)}$   & $-$16.2 & $-$16.3&$-$16.3&$-$16.3&$-$16.2 & $-$16.2\\
		$K\,\text{(MeV)}$  &282 & 281&281&281&272 & 245\\
		$E_\mathrm{sym}\,\text{(MeV)}$  &31.4&31.4&31.4 & 36.9&37.4 & 31.6 \\
		$L\,\text{(MeV)}$ &40 &40 &40 &111 & 118 & 47 \\
		$M^{*}/M$  & 0.794 &0.793 & 0.634& 0.634 & 0.595 & 0.761 \\
		\hline
	\end{tabular}
\end{center}
\end{table*}

\begin{table}[htbp]
\begin{flushleft}
\caption{Coupling constants and $\sigma$ meson mass for the TM1-family models. 
The $\omega$ and $\rho$ meson masses are fixed at 783 MeV and 770 MeV, respectively.}
	\label{parameter} 
	\begin{tabular}{>{\centering\arraybackslash}p{1.5cm} >{\centering\arraybackslash}p{1.4cm} >{\centering\arraybackslash}p{1.4cm} >{\centering\arraybackslash}p{1.4cm}
	>{\centering\arraybackslash}p{1.4cm}}
		\hline\hline
		& \textbf{TM1m*} & \textbf{TM1m} & \textbf{TM1e} & \textbf{TM1} \\
		\hline
		$g_{\sigma}$  & 7.15454 & 7.9353 & 10.0289 & 10.0289  \\
		$g_{\omega}$  & 8.59221 & 8.6317 & 12.6139 & 12.6139 \\
		$g_{\rho}$   & 11.52164 &11.5130 & 13.9714 &  9.2644  \\
		$g_{2}$ ($\text{fm}^{-1}$) & -7.86677 & -11.5163  & -7.2325 & -7.2325 \\
		$g_{3}$  & 40.55692 & 54.8872 & 0.6183 & 0.6183  \\
		$c_{3}$  & 0 & 0.00025 & 71.3075 & 71.3075   \\
		$\Lambda_{\text{v}}$  & 0.09465 & 0.0933 & 0.0429 & 0  \\
		$m_{\sigma}\,\text{(MeV)}$ & 463.680 & 511.198 & 511.198 & 511.198  \\
		\hline
	\end{tabular}
\end{flushleft}
\end{table}

\begin{figure}[htb]
	\centering
	\includegraphics[width=8 cm]{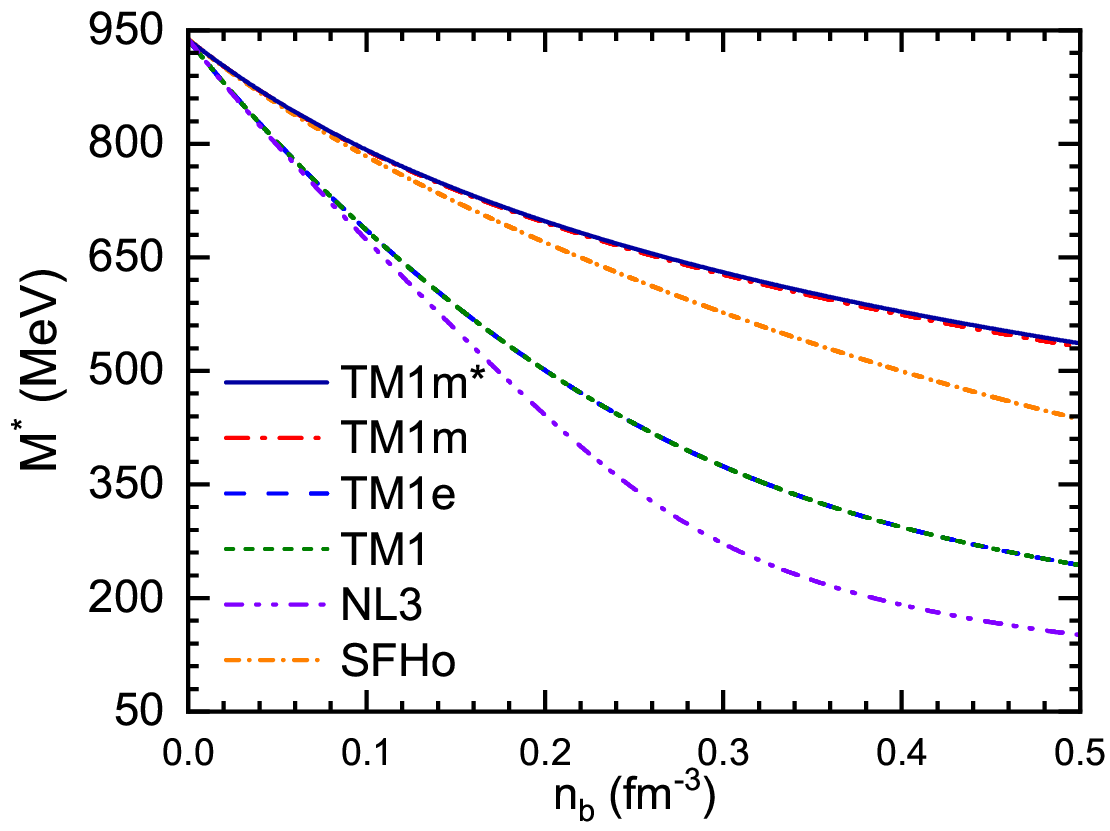}
	\caption{Effective nucleon mass $M^{\ast}$ as a function of baryon number density $n_b$ 
		for the TM1-family, NL3, and SFHo models.}
	\label{mn}
\end{figure}

The effective nucleon mass $M^{\ast}$ as a function of baryon number density $n_{b}$ is shown in Figure~\ref{mn}.
At low densities, the effective nucleon masses fall into two groups. TM1m*, TM1m, and SFHo have nearly 
identical values, while the other three models have similar but lower values.
As density increases, these values gradually diverge. 
At high densities, they are ranked from highest to lowest as follows: 
TM1m*, TM1m, SFHo, TM1e (equivalent to TM1), and NL3. 
The equivalence of TM1e and TM1 arises because they differ only in their isovector couplings.

\begin{figure}[htb]
\centering
\includegraphics[width=8 cm]{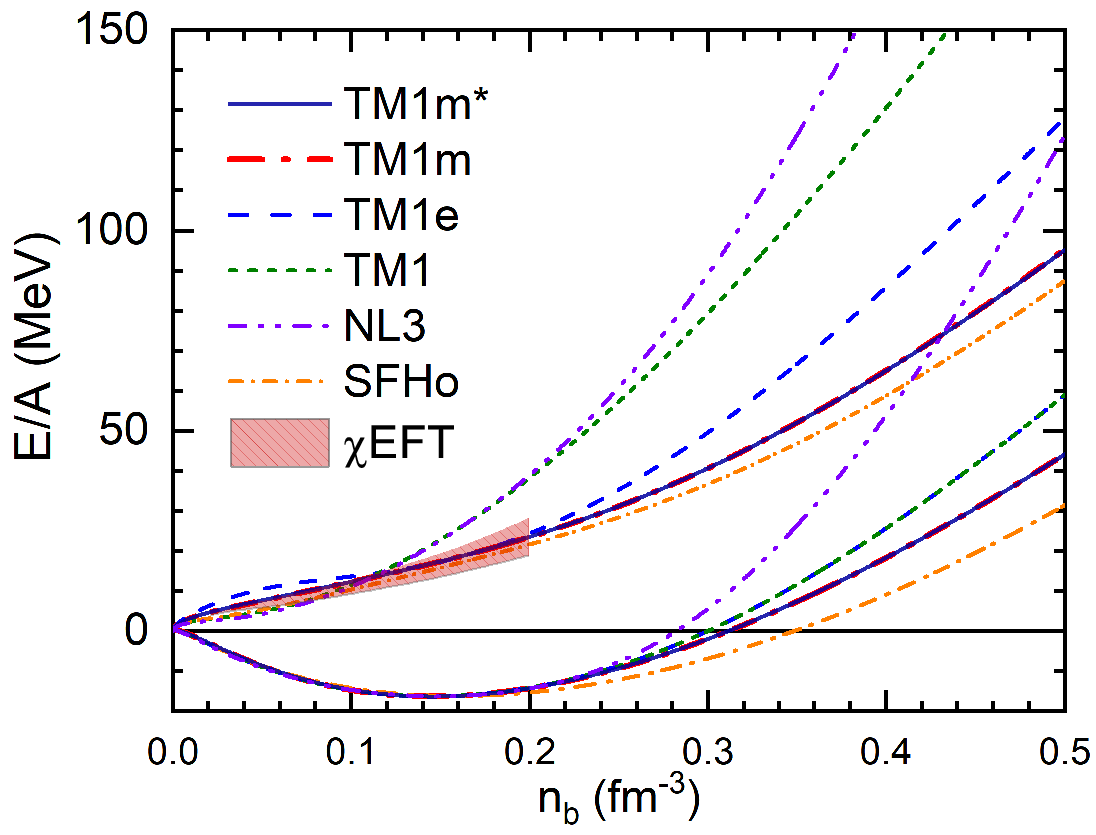}
\caption{Energy per nucleon $E/A$ of pure neutron matter and symmetric nuclear matter
as a function of baryon number density $n_b$. The red shaded band denotes the $E/A$ of pure neutron 
matter obtained from $\chi$EFT calculations~\citep{dris16} for comparison.}
	\label{ea}
\end{figure}

In Figure~\ref{ea}, we plot the energy per nucleon $E/A$ of pure neutron matter and symmetric nuclear 
matter as a function of baryon number density $n_b$.
For symmetric nuclear matter, all models are nearly identical at low densities 
and are consistent with the empirical saturation values, 
while they gradually diverge at higher densities. 
The NL3 model exhibits the most rapid increase, whereas the SFHo model shows the slowest.
This behavior is mainly attributed to differences in the nonlinear terms.
The curves for TM1e and TM1 lie between those for NL3 and SFHo and nearly coincide,
because they share the same isoscalar couplings. 
Similarly, the curves for TM1m* and TM1m are almost identical.
For pure neutron matter, the values of $E/A$ are determined not only by the isoscalar 
couplings but also by the isovector ones. 
At low densities, models with a smaller $L$ tend to yield higher $E/A$, 
while those with a larger $L$ result in lower $E/A$. 
At high densities, the correlation between $E/A$ and $L$ exhibits the opposite trend. 
It is shown that there are obvious discrepancies between the TM1e and TM1 curves for pure neutron matter,
in contrast to their nearly identical curves for symmetric nuclear matter.
Meanwhile, the TM1m* and TM1m curves almost coincide for both pure neutron matter 
and symmetric nuclear matter.
In addition, we compare the results of the RMF models considered here
with the $\chi$EFT uncertainty band for pure neutron matter~\citep{dris16,horn18}. 
It is seen that the TM1m and TM1m* models are compatible with the $\chi$EFT constraints
for densities up to $0.2\,\mathrm{fm}^{-3}$, whereas the other models lie more or less outside
the $\chi$EFT band. 

\begin{figure}[htb]
	\centering
	\includegraphics[width=8 cm]{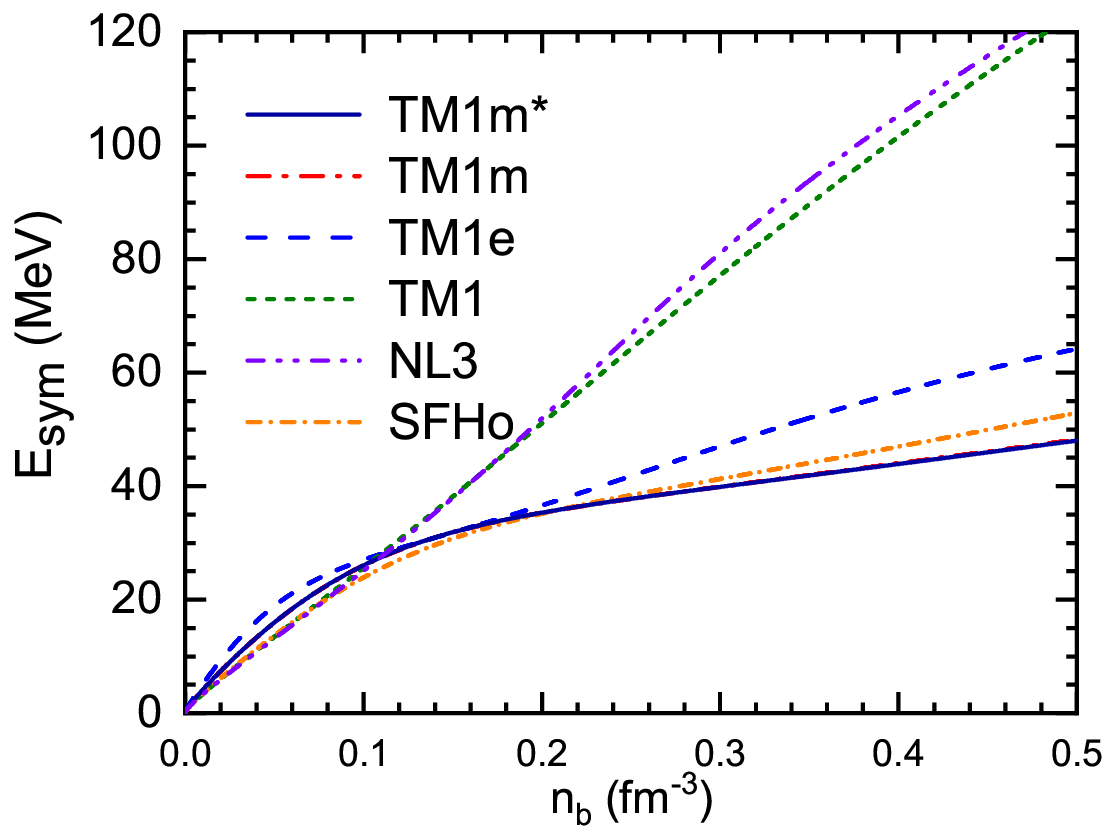}
	\caption{Same as Figure~\ref{mn}, but for symmetry energy $E_{\mathrm{sym}}$. }
	\label{esym}
\end{figure}

In Figure~\ref{esym}, we display the symmetry energy $E_{\mathrm{sym}}$ as a function of baryon number 
density $n_b$ for the six models considered in this work. To understand the properties of pasta structures 
from the $E_{\mathrm{sym}}$ curves, we discuss primarily the symmetry energy below saturation density. 
At $n_b=0.11\,\text{fm}^{-3}$, the TM1-family and NL3 models have almost identical 
values of $E_{\mathrm{sym}}$, ensuring reasonable predictions for the binding energies of finite nuclei. 
However, due to their different symmetry energy slopes $L$, these models exhibit 
divergent $E_{\mathrm{sym}}$ curves. Specifically, a larger $L$ leads to smaller $E_{\mathrm{sym}}$ 
at lower densities and larger $E_{\mathrm{sym}}$ at higher densities.
The opposite trend is observed for smaller $L$. 
The TM1m* and TM1m share the same $E_{\mathrm{sym}}$ and $L$ at saturation density, 
but at low densities their $E_{\mathrm{sym}}$ values are smaller than those of TM1e due to their larger $M^{\ast}$. This phenomenon is consistent with Figure~2 of \citet{choi21}.
We can understand this difference through the symmetry energy expression,
\begin{equation}
	E_{\mathrm{sym}} =\frac{k_F^2}{6 \sqrt{k_F^2+M^{\ast 2}}}
	+\frac{g_\rho^2 n_b}{8\left(m_\rho^2+2 \Lambda_{\mathrm{v}} g_\rho^2 g_\omega^2 \omega^2\right)}.
\end{equation}
It is shown that a larger $M^\ast$ suppresses the first term and yields smaller 
$E_{\mathrm{sym}}$ at lower densities. 
However, around the saturation density, this suppression is compensated by an increase of the second term,
resulting in identical values of $E_{\mathrm{sym}}$ for the TM1m*, TM1m, and TM1e models.
At higher densities, the influence of the effective nucleon mass becomes pronounced, 
leading to a greater divergence in the $E_{\mathrm{sym}}$ curves among these three models.
The SFHo model predicts $E_{\mathrm{sym}}$ comparable to those of TM1m* and TM1m 
at saturation density. However, at lower densities, the $E_{\mathrm{sym}}$ of SFHo 
is smaller than that of other models.

\section{Results and discussion} \label{sec:3}

In this section, we present numerical results based on the TM1-family, NL3, and SFHo models. 
First, we investigate the impact of effective nucleon mass and nuclear symmetry energy
on pasta phases occurring in the inner crust of neutron stars. Next, using the unified EOSs
based on these models, we explore macroscopic properties of neutron stars and analyze the
model dependence of the results. In the subsequent figures, we present the results of SFHo 
following those of TM1m* and TM1m, to facilitate comparison of effective nucleon mass effects.

\begin{figure}[htb]
	\centering
	\includegraphics[width=9 cm]{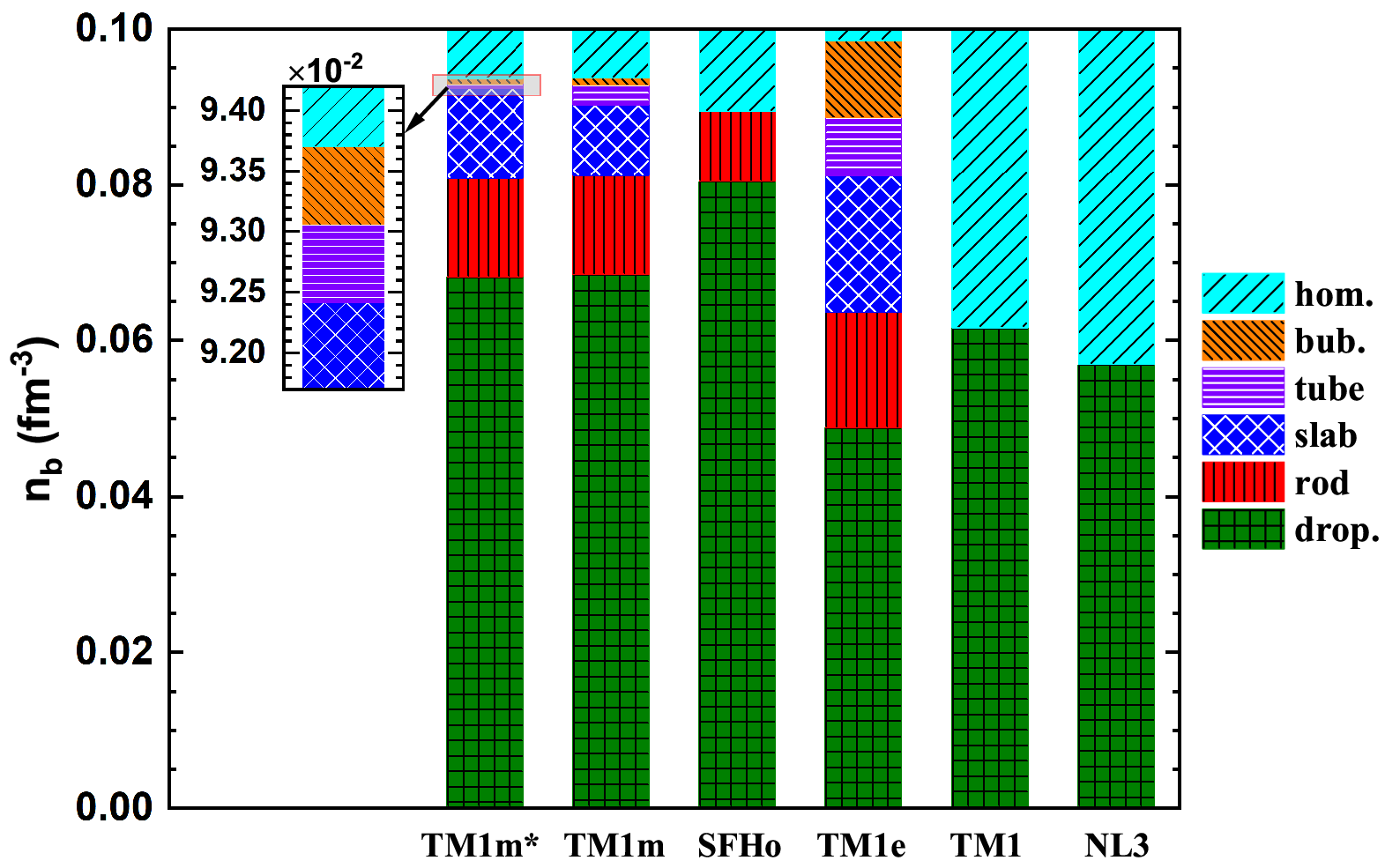}
	\caption{Phase diagram for the TM1-family, NL3, and SFHo models. The legend indicates the distinct 
		colors corresponding to the droplet, rod, slab, tube, bubble, and homogeneous phases. }
	\label{pasta_phase}
\end{figure} 

In Figure~\ref{pasta_phase}, we present pasta phase diagrams for the six models under consideration. 
Each pasta phase is depicted as a distinct colored region, with the vertical axis denoting 
the corresponding density range.
The TM1m* and TM1m models exhibit a similarly rich variety of nuclear pasta structures.
At densities below $\sim 0.09\,\text{fm}^{-3}$, both TM1m* and TM1m predict droplet, rod, 
and slab structures over similar density ranges. 
As the density increases, both models further develop tube and bubble structures 
within a very narrow density interval before transitioning to the homogeneous phase. 
The precise onset densities for the nonspherical pasta structures and homogeneous matter are listed 
in Table~\ref{Onset_densities}.
In addition, the TM1e model exhibits various pasta structures, with nonspherical configurations
appearing at a relatively low density of about $0.049\,\text{fm}^{-3}$. 
This phenomenon can be explained by its symmetry energy behavior at subsaturation densities.
According to the analysis within the liquid-drop model, 
the volume fraction of the liquid phase, $u$, plays a key role in the emergence of 
nonspherical configurations.
It has been reported in~\citet{peth95,oyam07} that
a spherical liquid drop becomes unstable against quadrupole deformations 
when the volume fraction $u$ reaches the fission-instability criterion, 
$u=\left(r_d / r_{\mathrm{ws}}\right)^3=1/8$, leading to the emergence of nonspherical configurations.
Furthermore, when $u > 1/2$, density-inverted pasta phases, such as tube and bubble configurations,
become energetically more favorable until the system transitions to uniform matter. 
Within an extended Thomas--Fermi approach using several Skyrme parametrizations~\citep{shch25}, 
the correlation between the occurrence of pasta phases
and the filling fraction $u$ has been extensively discussed, 
and the conclusion of~\citet{shch25} is consistent with the predictions of the liquid-drop model.
Generally, a larger $E_{\mathrm{sym}}$ leads to a higher $\mu_e$ based on the relation 
$\mu_e = \mu_n-\mu_p \simeq 4 \delta E_{\mathrm{sym}}$, where $\delta=1-2Y^L_p$ denotes 
the neutron excess with $Y^L_p$ being the proton fraction in the liquid phase. 
Consequently, a larger $\mu_e$ implies a larger $n_e$ and thus results in a larger volume 
fraction $u$ according to the relation 
$n_e = n_p^{L} u$ in the liquid-drop model~\citep{bao14b}.
Therefore, in the TM1e model, the larger symmetry energy $E_{\mathrm{sym}}$ at subsaturation 
densities leads to a larger volume fraction $u$, thereby inducing an earlier appearance of 
nonspherical configurations.
The SFHo model predicts nonspherical structures only above $0.08\,\text{fm}^{-3}$, 
a higher onset density than that of TM1m*, TM1m, and TM1e. This is mainly attributed to its 
lower $E_{\mathrm{sym}}$ at subsaturation densities as shown in Figure~\ref{esym}.

The transition densities to uniform matter exhibit model dependence. For TM1m* and TM1m cases,
the crust-core transition occurs at approximately $0.094 \, \text{fm}^{-3}$, which is slightly 
lower than that for TM1e. This can be explained by the symmetry energy curve as well. 
The smaller $E_{\mathrm{sym}}$ of TM1m* and TM1m at subsaturation densities leads to a higher 
neutron abundance and promotes neutron dripping from nuclei. Consequently, the dripped neutrons 
reduce the density contrast between the dense liquid phase and dilute gas phase, 
resulting in an earlier crust-core transition. 
In addition, the crust-core transition density in the SFHo model is about $0.0895 \, \text{fm}^{-3}$,
which is slightly lower than those of TM1m* and TM1m.
However, in the TM1 and NL3 models, the crust-core transition occurs at a significantly lower density
than in the other cases. Meanwhile, the transition proceeds directly from the droplet phase to
homogeneous matter without passing through nonspherical phases. This phenomenon can be attributed to
the larger symmetry energy slope $L$ in TM1 and NL3. 
As discussed in~\citet{duco11} and~\citet{bao15}, the crust-core transition density is negatively 
correlated with $L$. Moreover, it was reported in~\citet{Gear11} that nonspherical configurations 
disappear at the upper end of the experimentally constrained range of $L$, which is consistent 
with the behavior observed in the TM1 and NL3 models.

\begin{table}[htb]
	\begin{center}
		\caption{Onset densities of nonspherical structures and homogeneous matter 
			in the TM1-family, NL3, and SFHo models.}
		\label{Onset_densities}
		\setlength{\tabcolsep}{2.6mm}
		\begin{tabular}{lccccc} 
			\hline\hline
			Model & \multicolumn{5}{c}{Onset density (fm$^{-3}$)} \\
			\cline{2-6}
			& rod & slab & tube & bubble & hom.   \\
			\hline
			TM1m* & 0.0682 & 0.0809 & 0.0924 & 0.0931 & 0.0937 \\
			TM1m  & 0.0685 & 0.0812 & 0.0903 & 0.0929 & 0.0938 \\
			SFHo  & 0.0805 & $-$    & $-$    & $-$    & 0.0895 \\
			TM1e  & 0.0489 & 0.0636 & 0.0812 & 0.0887 & 0.0986 \\
			TM1   & $-$    & $-$    & $-$    & $-$    & 0.0616 \\
			NL3   & $-$    & $-$    & $-$    & $-$    & 0.0570 \\
			\hline
		\end{tabular}
	\end{center}
\end{table}

\begin{figure*}[htb]
	\centering
	\includegraphics[width=18 cm]{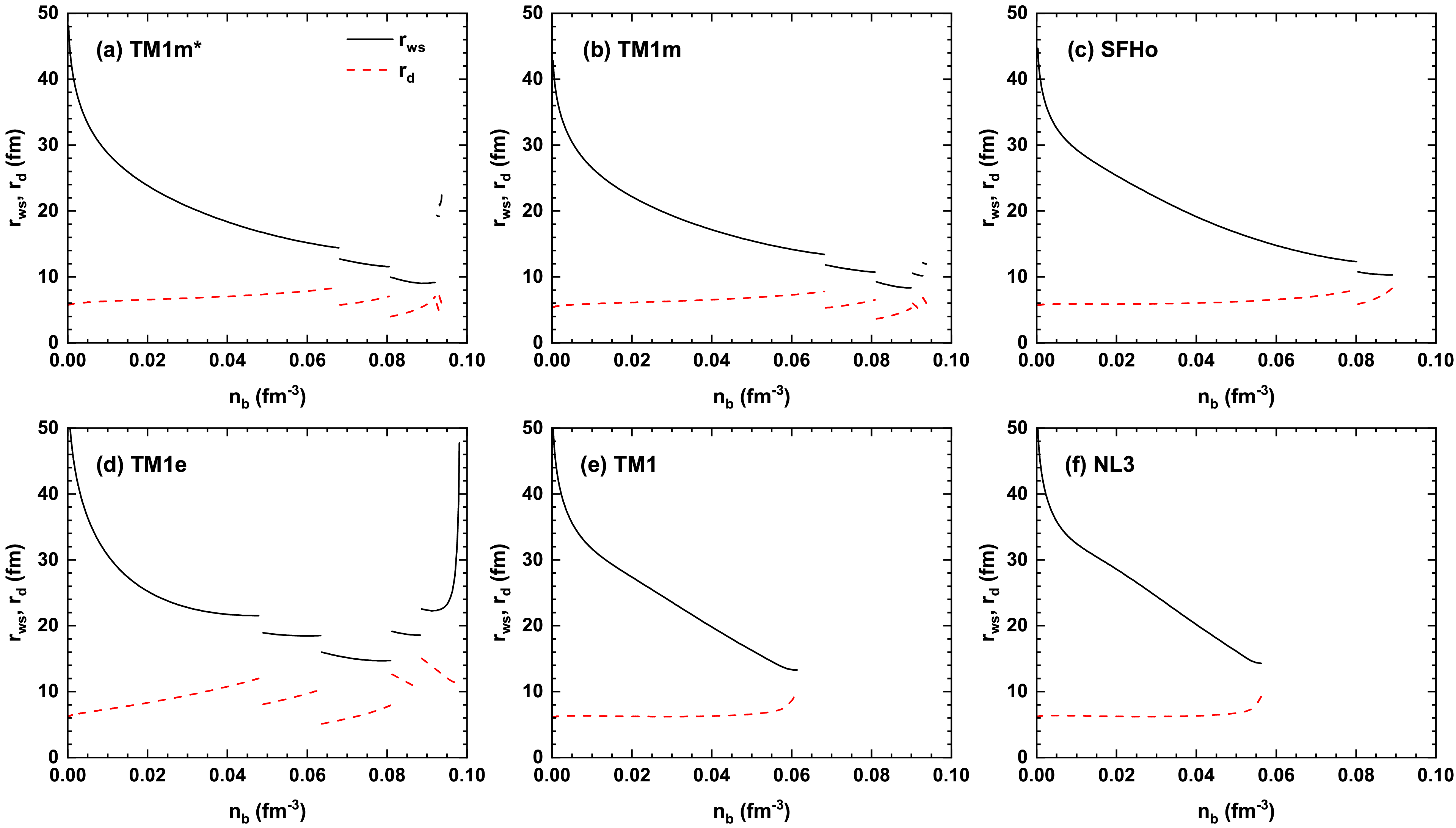}
	\caption{The size of the Wigner--Seitz cell $r_{\text{ws}}$ and that of the inner part $r_{d}$ 
		as a function of baryon number density $n_b$ for the TM1-family, NL3, and SFHo models.
		Discontinuities in the curves appear at the transition points between different
		pasta shapes.}
	\label{rwsrd}
\end{figure*}

In Figure~\ref{rwsrd}, we display the size of the Wigner--Seitz cell, $r_{\text{ws}}$, and that of 
the inner part, $r_{d}$, as defined in Equation~(\ref{rd}). In the droplet phase, $r_{\text{ws}}$ 
decreases and $r_{d}$ increases with increasing density. 
It is observed that $r_{\text{ws}}$ and $r_{d}$ exhibit pronounced discontinuities at the transition 
points between different pasta shapes. 
This phenomenon arises because the present calculations are restricted to 
a limited set of pasta shapes without allowing for intermediate configurations.
The trends of $r_{\text{ws}}$ and $r_{d}$ in TM1m*, TM1m, and SFHo are very similar.
On the other hand, the TM1e model reveals very complex behaviors for $r_{\text{ws}}$ and $r_{d}$,
including five pasta shapes. In the droplet phase, a pronounced increase in $r_{d}$
leads to a transition to a nonspherical phase. In the bubble phase prior to the crust-core 
transition, $r_{\text{ws}}$ increases rapidly while $r_{d}$ decreases. 
This implies that the liquid phase outside the bubble becomes increasingly dominant 
before the transition to uniform matter.
In contrast, the TM1 and NL3 models undergo a direct transition from the droplet phase to uniform matter 
without nonspherical phases. The trends of $r_{\text{ws}}$ and $r_{d}$ in TM1 and NL3
are similar to those in the droplet phase of other models. 
The larger values of $L$ in these two models result in an earlier transition to uniform matter.

\begin{figure*}[htb]
	\centering
	\begin{tabular}{ccc}
		\includegraphics[width=6 cm]{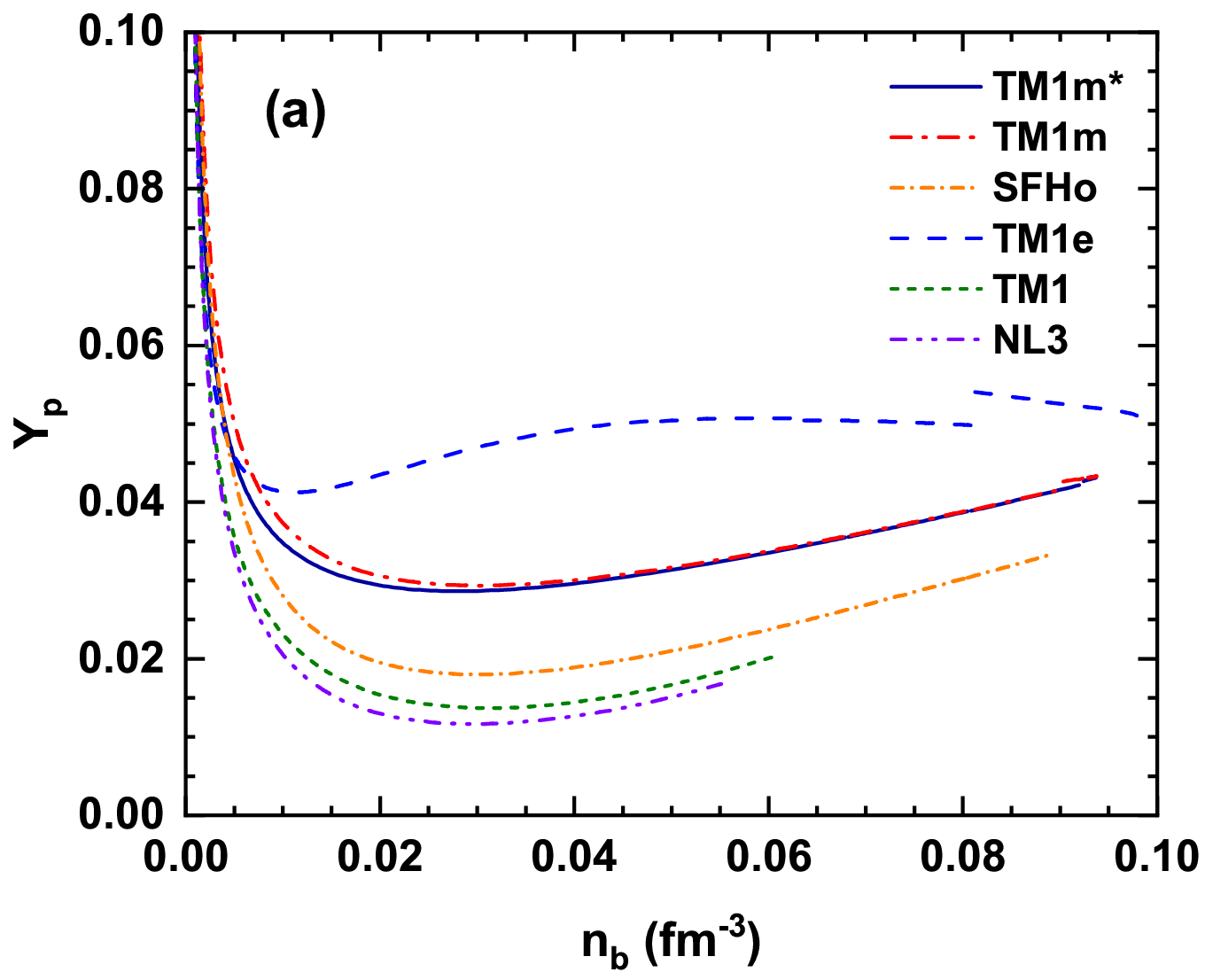}&
		\includegraphics[width=6 cm]{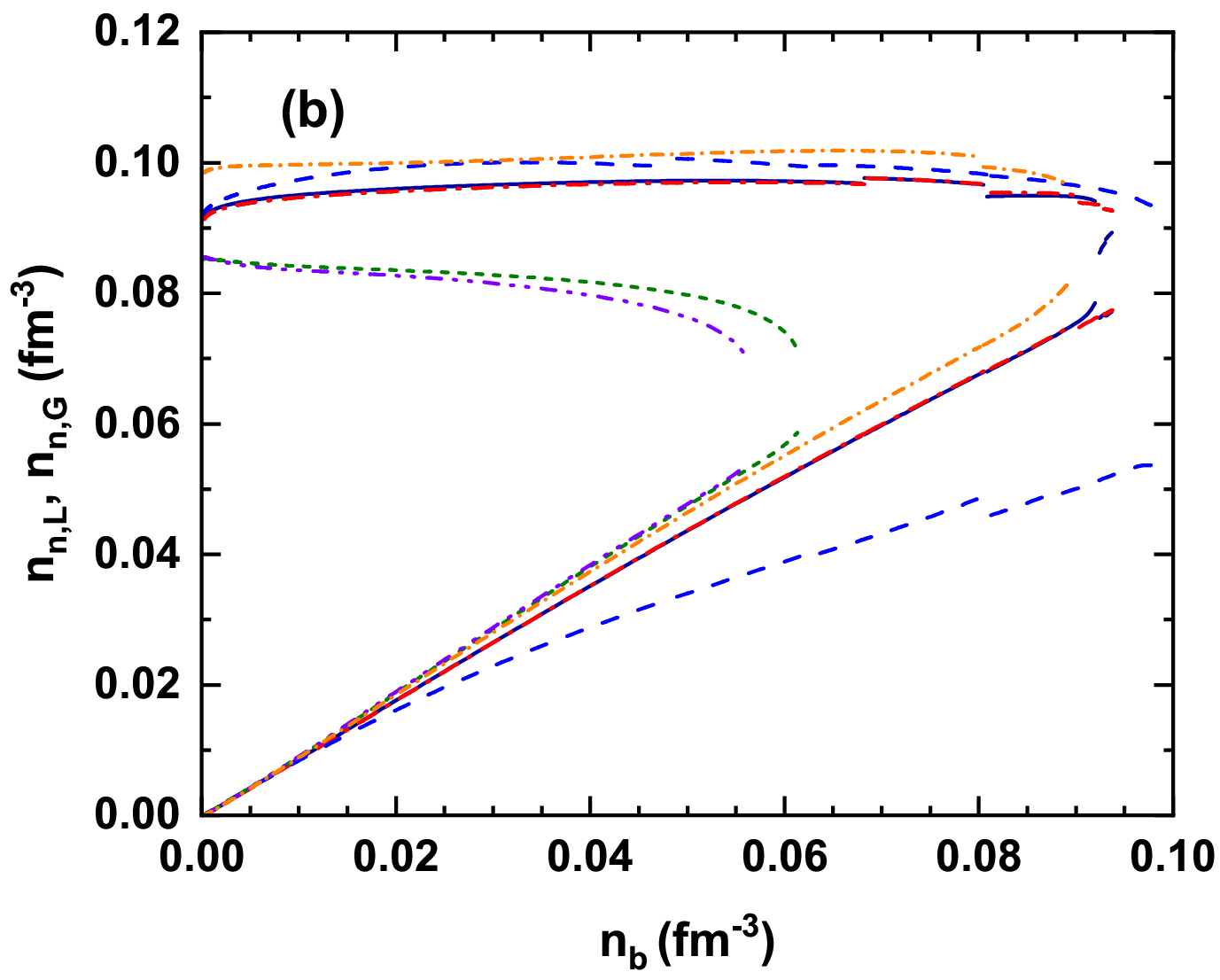}&
		\includegraphics[width=6 cm]{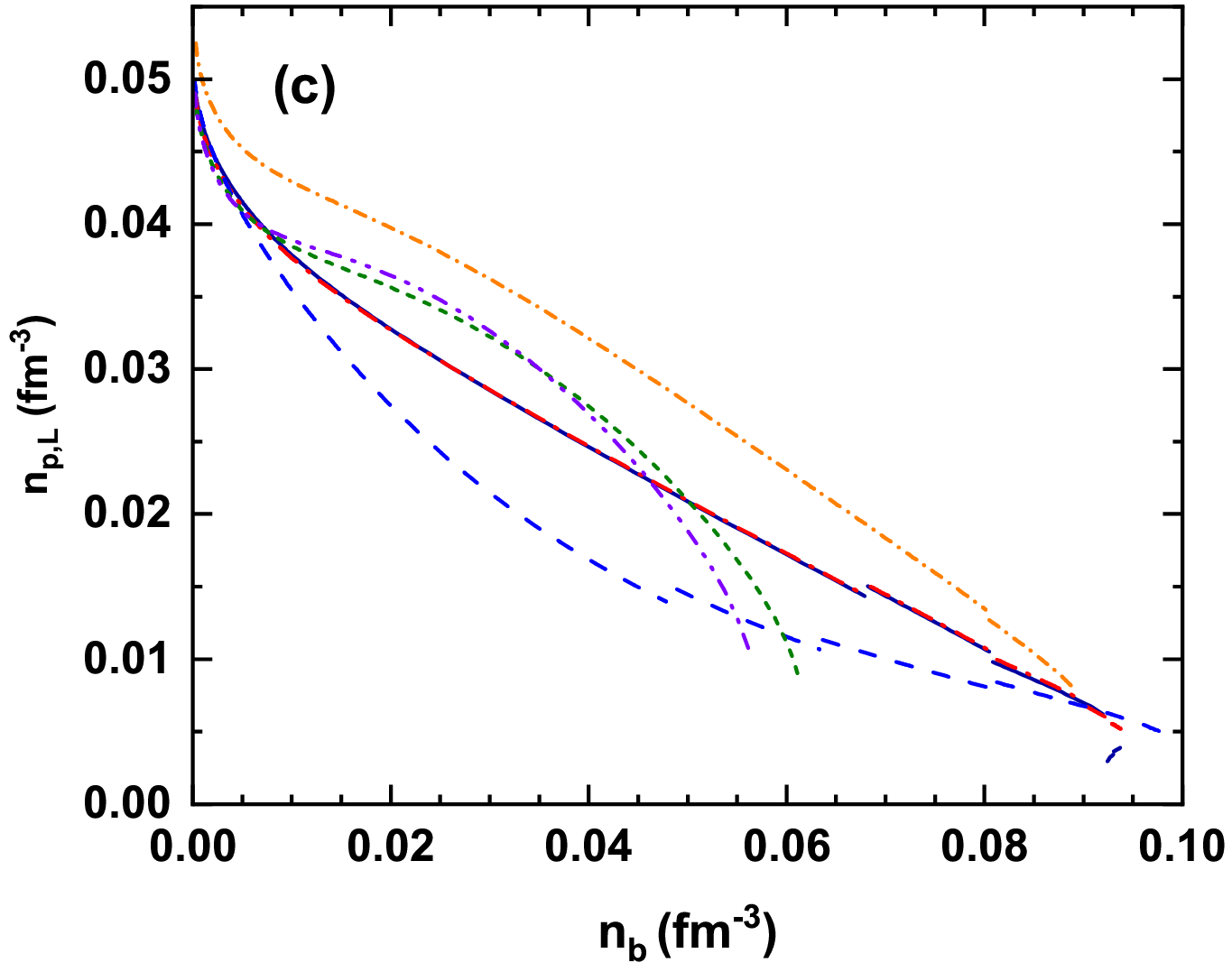}\\
	\end{tabular}
	\caption{Comparison of the average proton fraction $Y_p$ (a), the neutron densities of the liquid phase and the gas phase, $n_{\text{n,L}}$ and $n_{\text{n,G}}$, at the center or boundary of the cell (b), and the proton density of the liquid phase, $n_{\text{p,L}}$, at the center or boundary of the cell (c) among the TM1-family, NL3, and SFHo models.}
	\label{ypnnnp}
\end{figure*}

We explore the model dependence of pasta phase properties calculated within the TF framework.
In Figure~\ref{ypnnnp}, we show the following quantities: 
(a) the average proton fraction $Y_p$ in the Wigner--Seitz cell; 
(b) the neutron densities of the liquid phase and the gas phase, $n_{\text{n,L}}$ and $n_{\text{n,G}}$, 
defined as the density at the center or boundary of the cell; 
and (c) the proton density of the liquid phase, $n_{\text{p,L}}$, at the center or boundary of the cell.
As can be seen in Figure~\ref{ypnnnp}(a), the average proton fraction $Y_p$ in all models exhibits 
a rapid decrease at low densities and then varies gradually as the density increases.
In particular, the $Y_p$ curve for the TM1e model is obviously higher than those of other models,
which can be attributed to its larger $E_{\rm sym}$ at low densities, as displayed in Figure~\ref{esym}.
It is well known that a larger $E_{\rm sym}$ favors a higher $Y_p$ in homogeneous $\beta$-equilibrium matter.
The curves for TM1m* and TM1m are very close to each other, while those of SFHo lie lower.
The TM1 and NL3 models predict even lower $Y_p$ owing to their smaller $E_{\rm sym}$ at low densities. 
In Figure~\ref{ypnnnp}(b), the curve of $n_{\text{n,G}}$ in the TM1e model is significantly lower 
than those in other models.
On the other hand, the curves of $n_{\text{n,L}}$ in the TM1 and NL3 models are remarkably lower,
owing to their larger values of $L$. 
It has been reported in \citet{oyam07}, \citet{gril12} and \citet{bao15} that
a larger $L$ favors a smaller $n_{\text{n,L}}$ and a larger $n_{\text{n,G}}$ in nuclear pasta phases.
In Figure~\ref{ypnnnp}(c), the behavior of $n_\text{{p,L}}$ is somewhat complicated. 
In general, $n_{\text{p,L}}$ decreases with increasing density.
The model dependence of the $n_{\text{p,L}}$ curves can be roughly explained
by the variation of chemical potentials (see Figure~\ref{munpe} below). 
A smaller $\mu_p$ in the TM1e model, as presented in Figure~\ref{munpe}(b), corresponds to a 
lower $n_{p,L}$ in Figure~\ref{ypnnnp}(c). 
The results of the TM1m* and TM1m models are nearly identical, which originates from their 
similar saturation properties.

We plot the chemical potentials of neutrons $\mu_n$, protons $\mu_p$, and electrons $\mu_e$ within 
the Wigner--Seitz cell in panels (a), (b), and (c) of Figure~\ref{munpe}, respectively. 
These chemical potentials are intensive quantities and play an important role in the TF calculation.
With increasing density, both $\mu_n$ and $\mu_e$ increase monotonically, whereas $\mu_p$ decreases.
The model dependence of chemical potentials can be explained by the behavior of symmetry energy.
As discussed above, a larger $E_{\mathrm{sym}}$ leads to a higher $\mu_e$ based on the relation 
$\mu_e = \mu_n-\mu_p \simeq 4 \delta E_{\mathrm{sym}}$. Therefore, for 
$n_b < 0.06\ \text{fm}^{-3}$, the TM1e model yields the highest $\mu_n$ and $\mu_e$, 
as well as the lowest $\mu_p$, owing to its relatively large $E_{\rm sym}$ at low densities.
Furthermore, the effective mass and the $\rho$ meson coupling can also affect the chemical potentials.
From Equations~(\ref{chem_pot_p}) and~(\ref{chem_pot_n}), we find that a larger $g_\rho$ results in 
a larger difference between $\mu_n$ and $\mu_p$. 
Compared with the TM1e model, the larger effective mass $M^{\ast}$ in the TM1m* and TM1m models 
suppresses the kinetic energy contribution to the chemical potential. 
Meanwhile, their smaller $g_\rho$ reduces the difference between $\mu_n$ and $\mu_p$. 
Consequently, the $\mu_n$ curves of TM1m* and TM1m 
lie below that of TM1e, as presented in Figure~\ref{munpe}(a).

\begin{figure*}[htb]
	\centering
	\begin{tabular}{ccc}
		\includegraphics[width=6 cm]{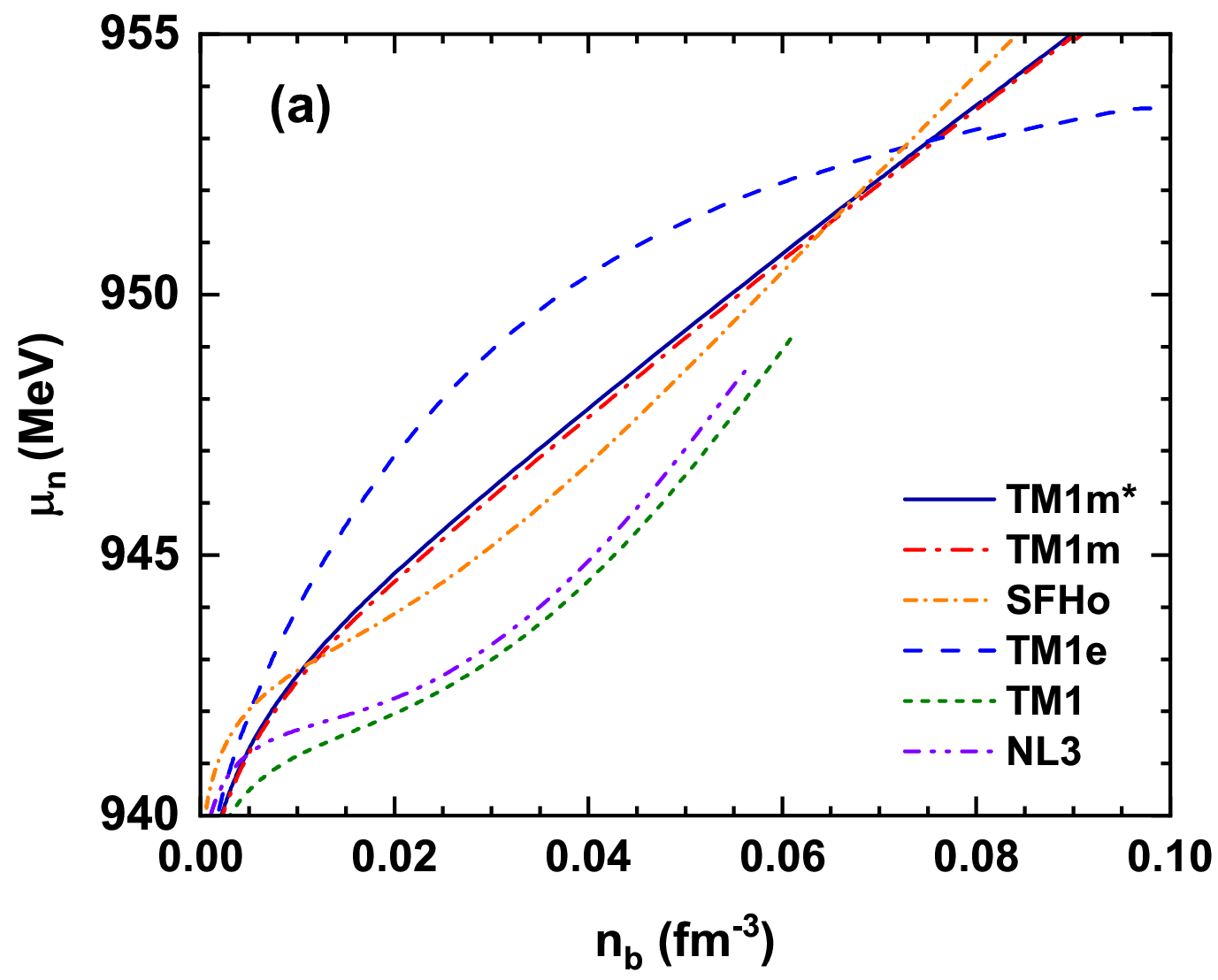}&
		\includegraphics[width=6 cm]{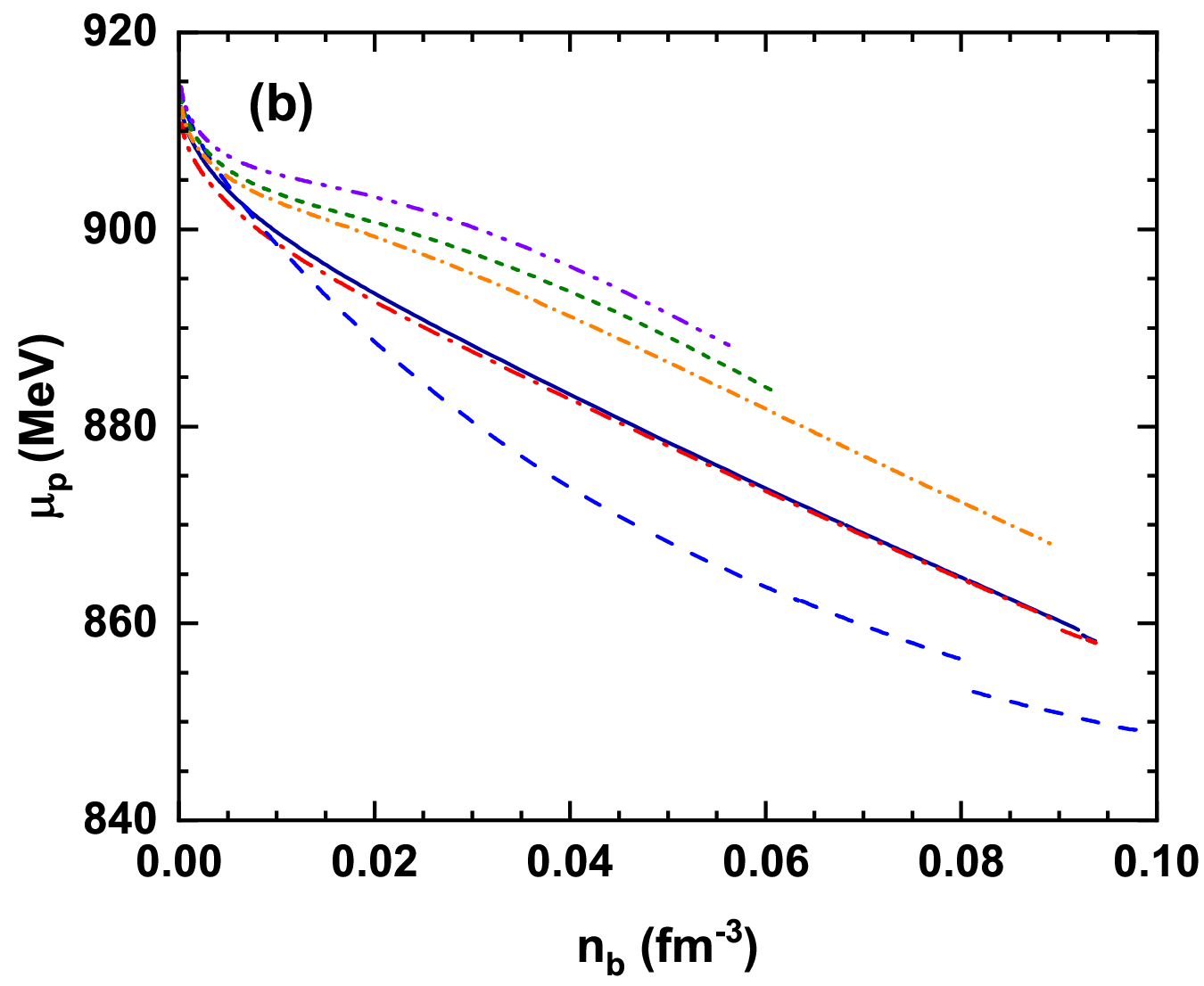}&
		\includegraphics[width=6 cm]{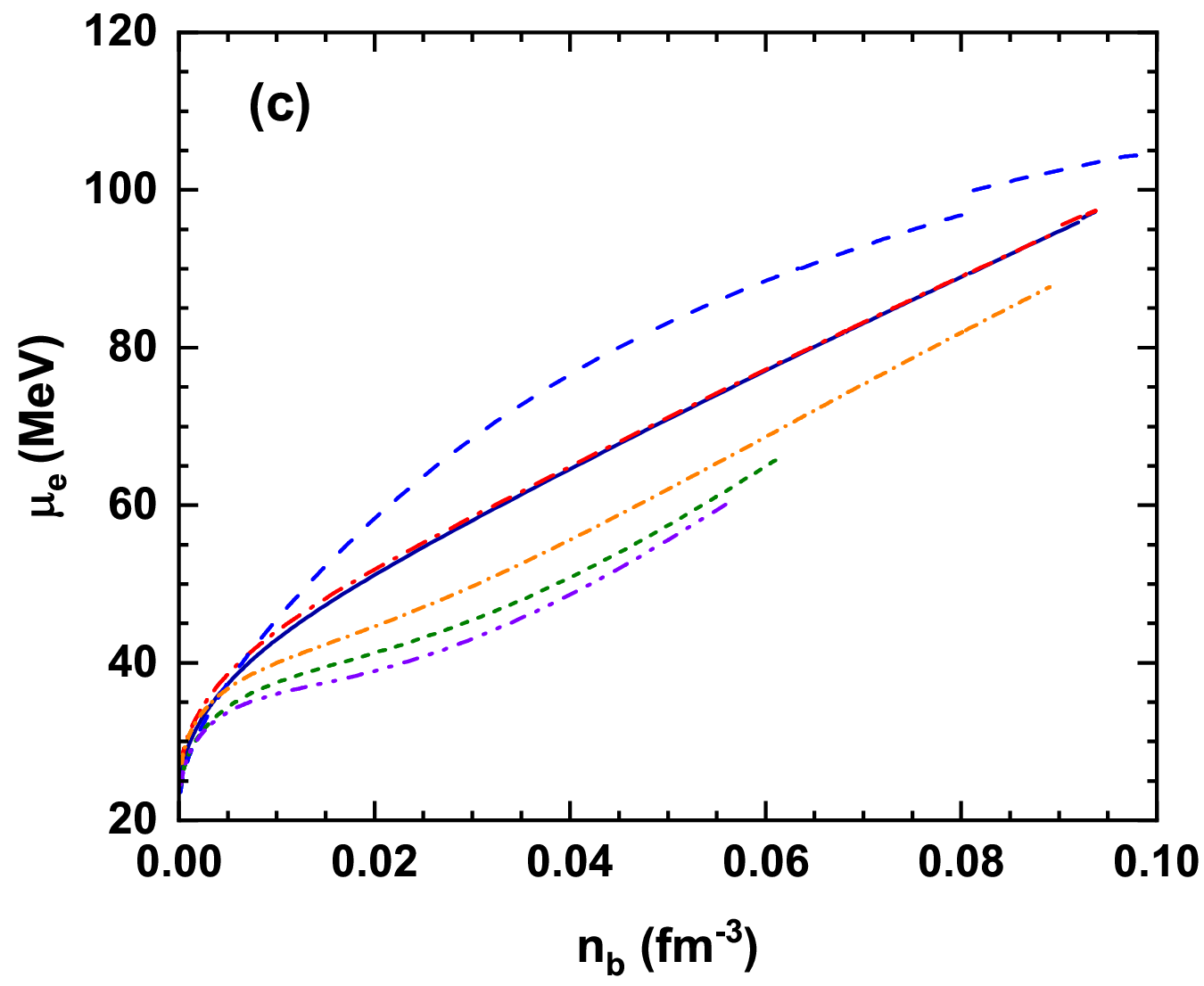}\\
	\end{tabular}
	\caption{Chemical potentials $\mu_n$, $\mu_p$, and $\mu_e$ as a function of baryon number density $n_b$. }
	\label{munpe}
\end{figure*}

\begin{figure}[htb]
	\centering
	\includegraphics[width=8.5 cm]{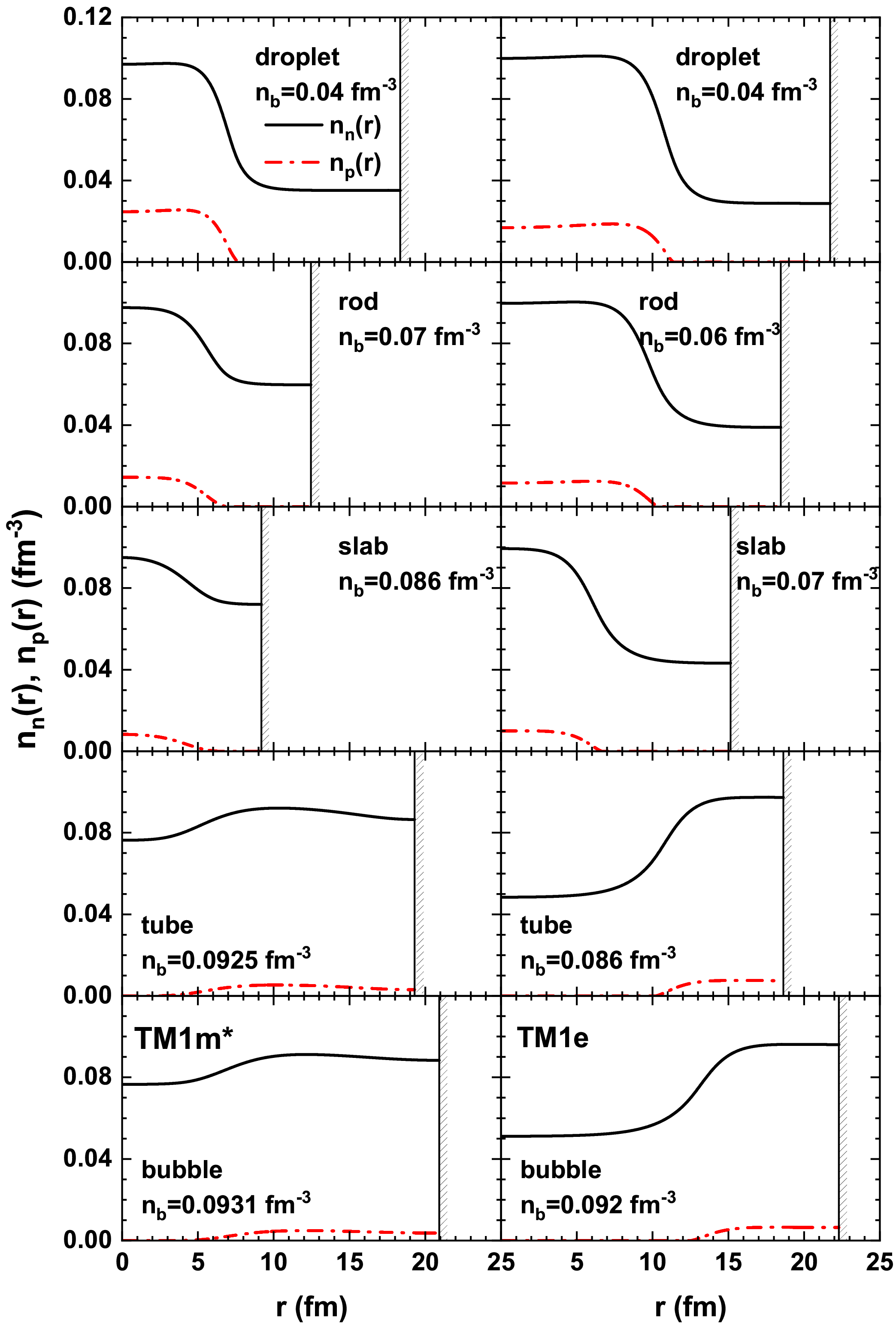} 
	\caption{Density distributions of neutrons (black solid lines) and protons (red dot-dashed lines) 
		in the Wigner--Seitz cell for TM1m* and TM1e. The cell boundary is indicated by hatching.}
	\label{dendis}
\end{figure}

To examine the influence of effective mass on pasta phase properties, we display in 
Figure~\ref{dendis} the nucleon density distributions for TM1m* and TM1e at typical densities.
From top to bottom, we present five cases corresponding to the droplet, rod, slab, tube, and bubble 
configurations, with TM1m* shown in the left panels and TM1e in the right panels.
In the top panels for the droplet phase at $n_b = 0.04\ \text{fm}^{-3}$, both models exhibit a clear 
separation between the liquid and gas phases. Compared with the TM1e model, the TM1m* model
yields distinctly smaller radii for the droplet and the Wigner--Seitz cell.
This is consistent with the results presented in Figure~\ref{rwsrd}.
Similar differences are also observed in the rod and slab phases.
As observed in the tube and bubble phases, the TM1m* model yields more diffuse density distributions
and a larger $n_{\text{n,G}}$, compared with the TM1e model. 
The difference in $n_{\text{n,G}}$ between TM1m* and TM1e is also evident in Figure~\ref{ypnnnp}(b).

\begin{figure}[htb]
	\centering
	\includegraphics[width=8 cm]{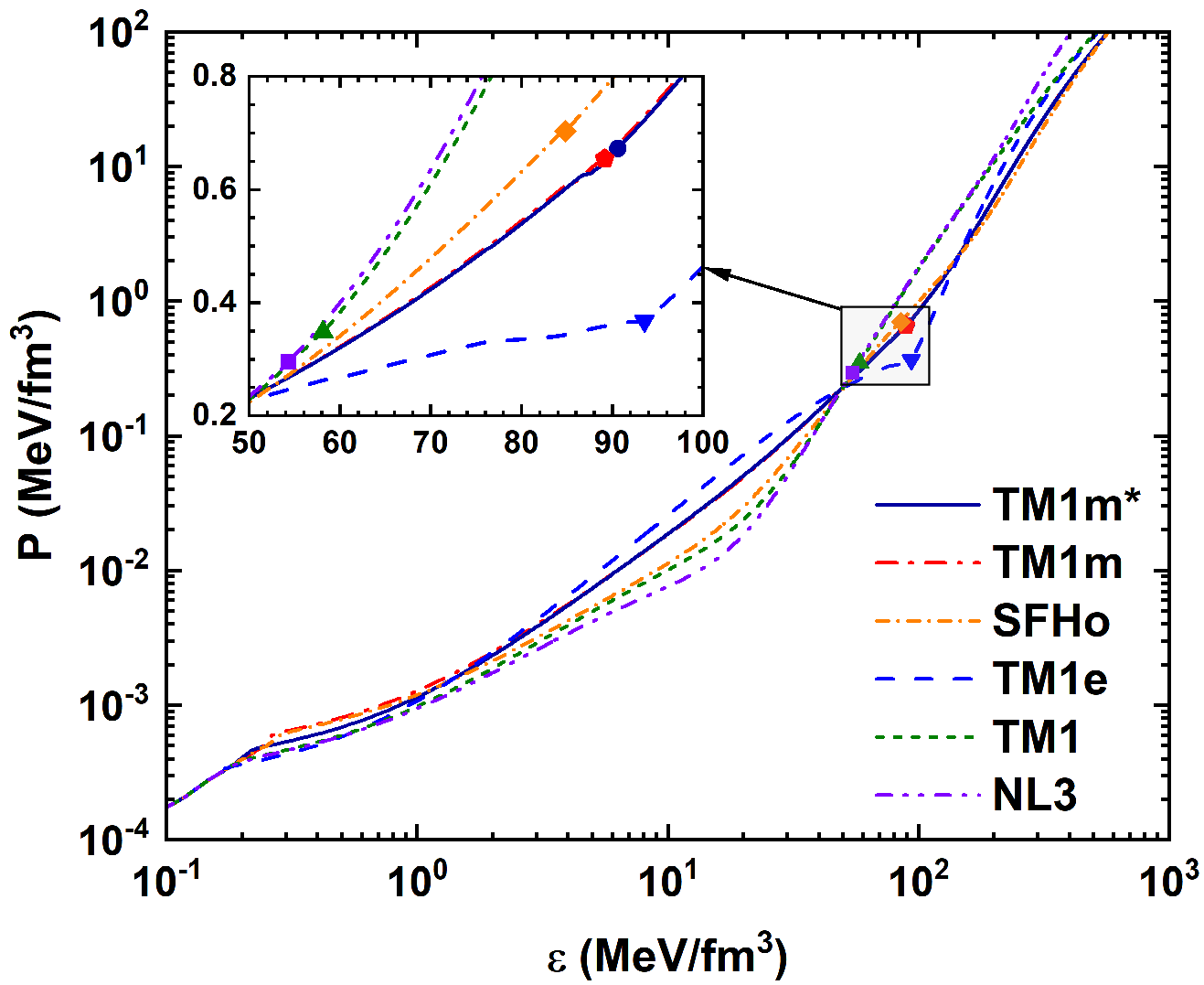}
	\caption{Pressure $P$ as a function of energy density $\varepsilon$ for the TM1-family, NL3, and SFHo 
		models. The crust-core transition is indicated by the filled symbols in the inset.}
	\label{eos}
\end{figure}

To obtain neutron-star properties, we solve the Tolman--Oppenheimer--Volkoff (TOV) equation
using the full EOS that covers a wide density range from the stellar surface to the dense core.
Generally, the EOS consists of three regions: the outer crust, the inner crust, and the liquid core. 
The outer crust, composed of neutron-rich nuclei embedded in an electron gas, plays only a minor role 
in determining macroscopic neutron-star properties such as the mass-radius relation. 
We adopt the well-known Baym--Pethick--Sutherland (BPS) EOS~\citep{baym71b} for the outer crust.
Furthermore, we construct a unified EOS by smoothly connecting the inner crust EOS obtained within
the TF approximation to the core EOS at the crust-core transition density, 
adopting the same RMF model for both the inner crust and core EOSs.
In Figure~\ref{eos}, we display the pressure $P$ as a function of energy density $\varepsilon$ obtained
in the TM1-family, NL3, and SFHo models. The crust-core transition is indicated by the filled symbols 
in the inset. It is well known that the pressure of pure neutron matter is mainly determined 
by the symmetry energy slope $L$. Therefore, the TM1 and NL3 models with larger slope values yield relatively
larger pressures at higher densities in the core region. In contrast, the TM1m*, TM1m, and SFHo models
result in smaller pressures at higher densities, which is mainly due to their slow increase of 
$E_{\mathrm{sym}}$ (see Figure~\ref{esym}), caused by a larger effective mass $M^{\ast}$.
Correspondingly, the TM1m*, TM1m, and SFHo models yield softer EOSs because the kinetic
contribution to the pressure is reduced due to a larger value of $M^{\ast}$.

\begin{figure}[htb]
	\centering
	\includegraphics[width=8 cm]{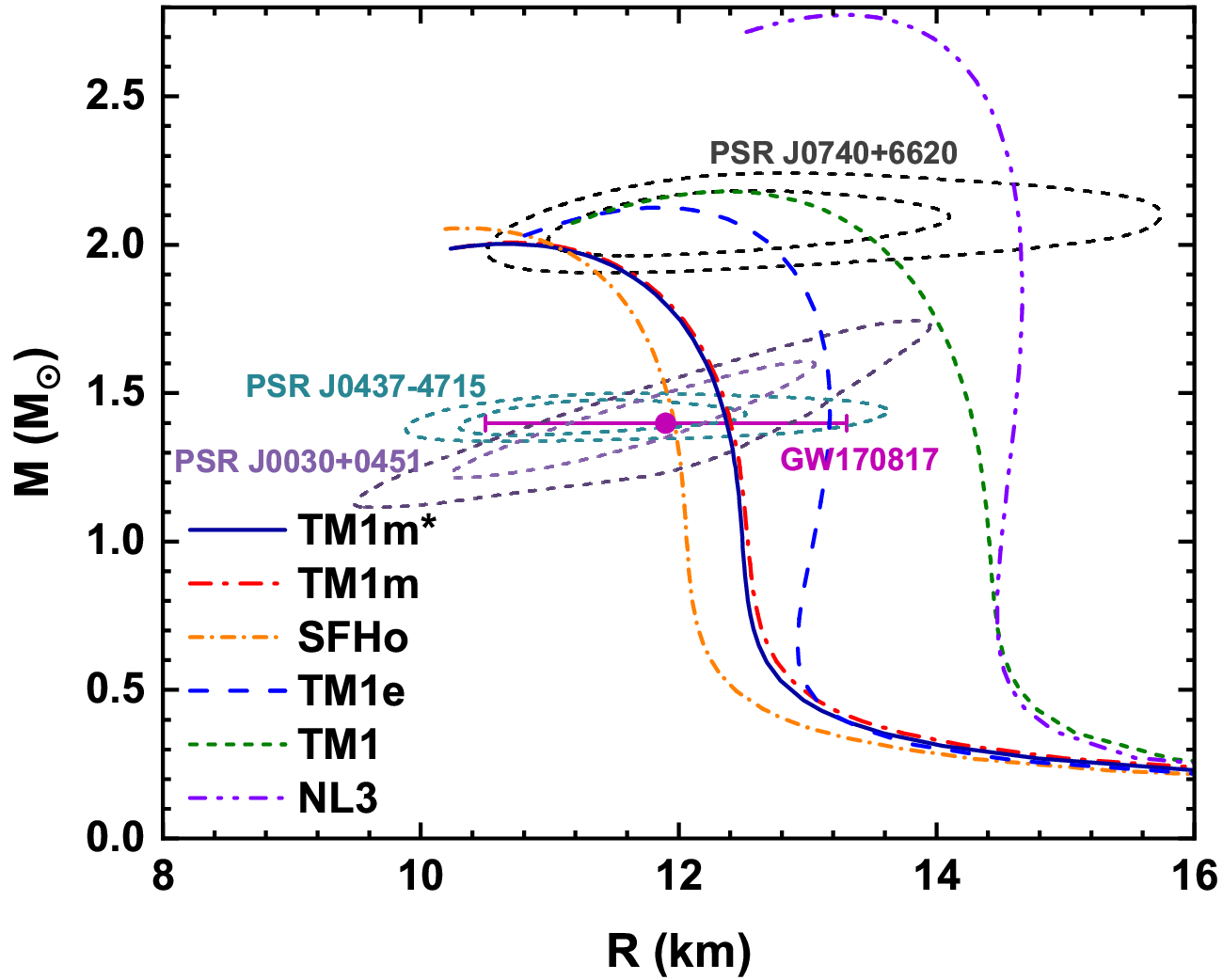}
	\caption{Mass-radius relations of neutron stars obtained in the TM1-family, NL3, and SFHo models. 
		The magenta horizontal line denotes the constraint on $R_{1.4}$ from GW170817~\citep{abbo18}.
		The short-dashed contours show the 68.3\% and 95.4\% credible mass-radius constraints 
		from PSR J0030+0451~\citep{vinc24}, PSR J0437-4715~\citep{chou24}, 
		and PSR J0740+6620~\citep{salm24}. 
	}
	\label{mr}
\end{figure}

With the unified EOSs obtained, we compute the mass‑radius relations of neutron stars by solving 
the TOV equation for the six models considered in this work. 
In Figure~\ref{mr}, the resulting mass‑radius relations are presented and compared with 
current astrophysical observations. 
It is shown that the TM1m* and TM1m models yield nearly identical results due to their similar
saturation properties, whereas the SFHo model predicts relatively small radii. 
The larger effective mass $M^{\ast}$ in these models suppresses the kinetic energy contribution 
to the pressure and leads to a softer EOS, which favors a smaller radius. 
Similar behavior was also reported in \citet{horn18}.
On the other hand, the TM1 and NL3 models produce much larger radii of neutron stars, owing to
their larger symmetry energy slope $L$. This is consistent with the positive correlation
between $L$ and $R_{1.4}$, where $R_{1.4}$ denotes the radius of a canonical $1.4 \, M_{\odot}$ neutron star. 
The $R_{1.4}$ results for TM1m*, TM1m, SFHo, and TM1e are compatible with observational
constraints from the gravitational wave event GW170817~\citep{abbo18}, as well as simultaneous
mass-radius measurements by NICER for PSR J0030+0451~\citep{vinc24}
and PSR J0437-4715~\citep{chou24}. 
The maximum neutron-star masses predicted by these models satisfy 
﻿the requirement of being larger than $2 \, M_{\odot}$. 
Meanwhile, their mass-radius relations are compatible with observational
constraints from NICER measurements of PSR J0740+6620~\citep{salm24}.

\begin{figure*}[htb]
	\centering
	\begin{tabular}{cc}
		\includegraphics[width=8 cm]{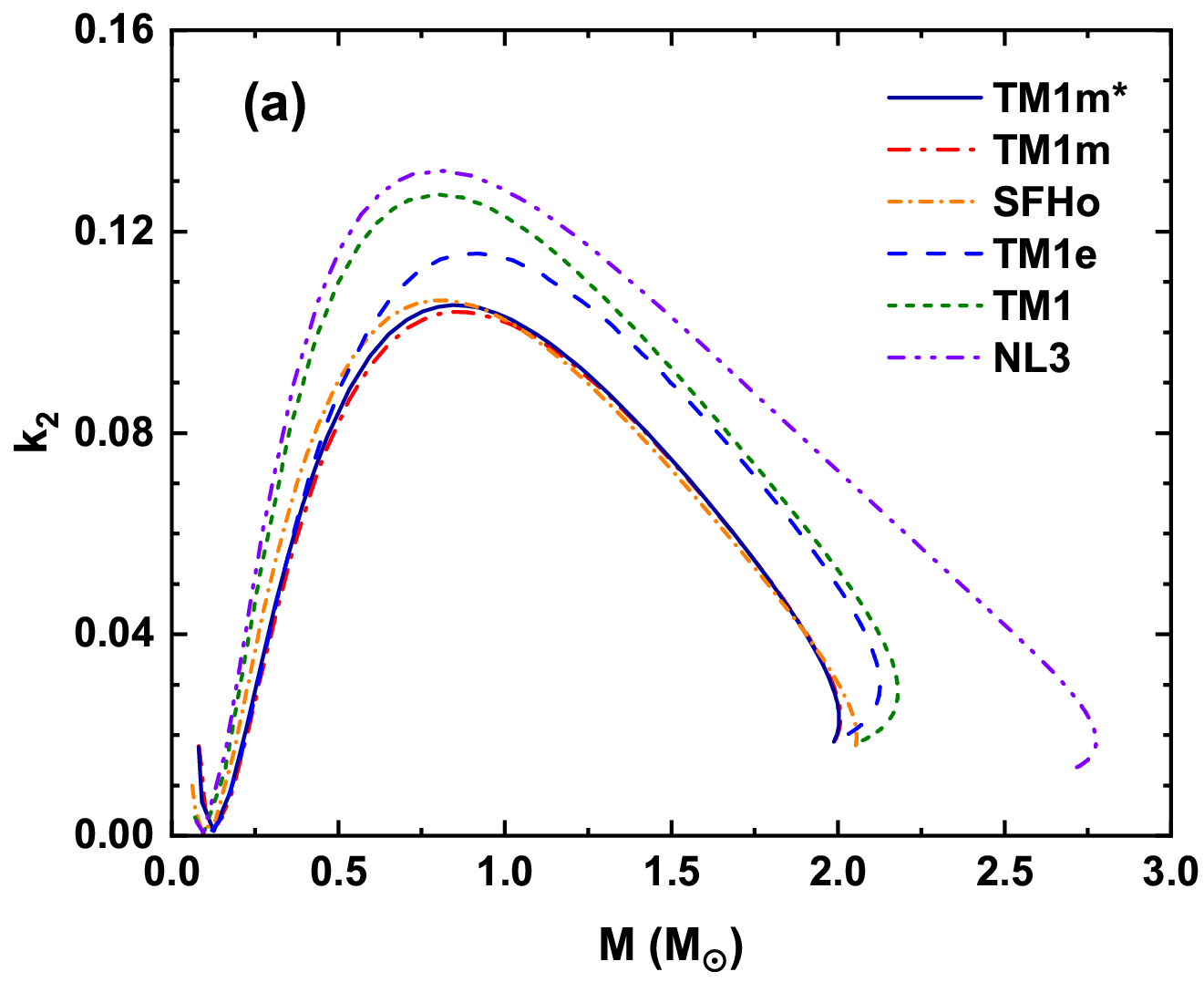}&
		\includegraphics[width=8 cm]{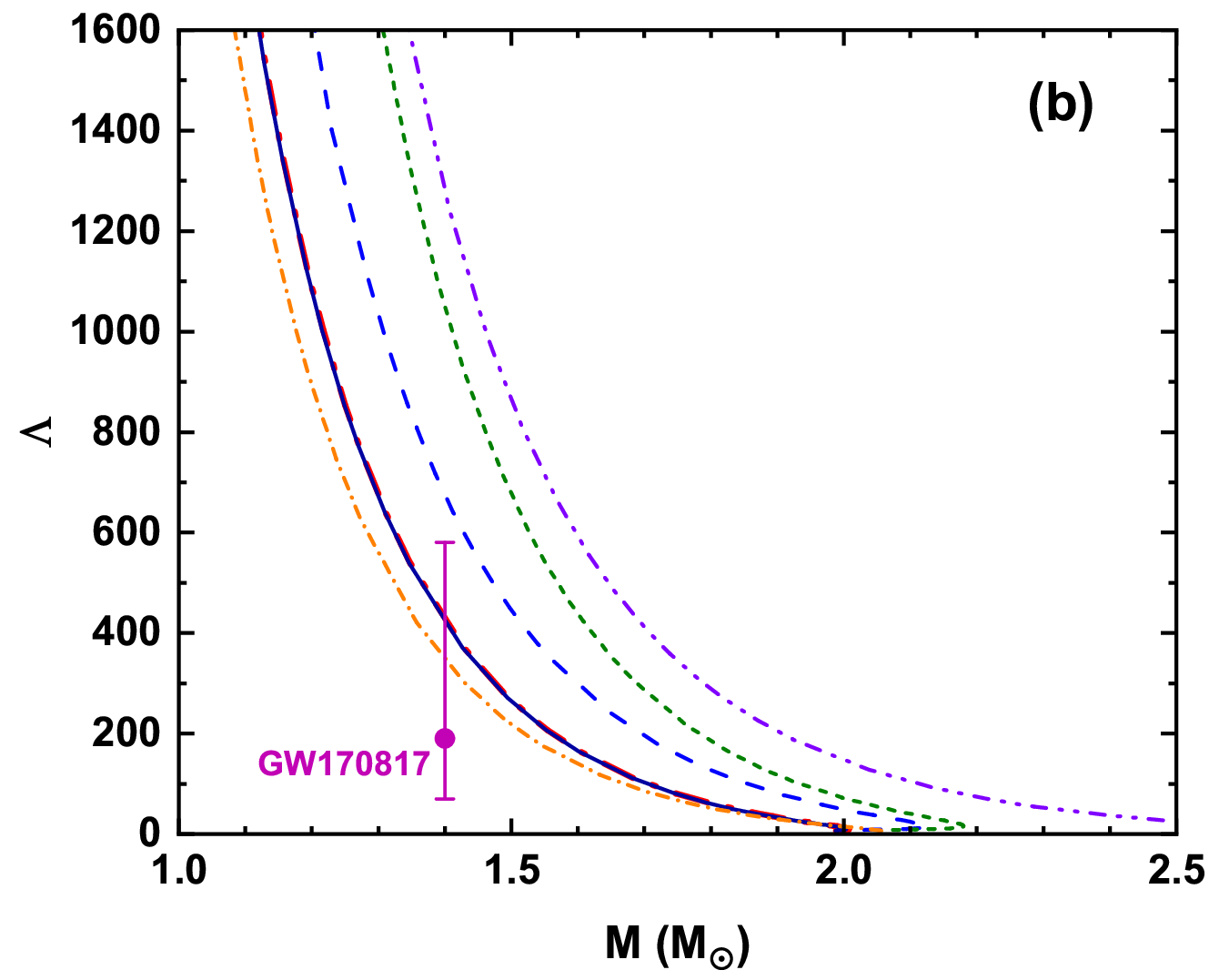}\\
	\end{tabular}
	\caption{Love number $k_2$ and dimensionless tidal deformability $\Lambda$ 
		as a function of the neutron star mass $M$. The magenta vertical line  denotes 
		the constraint on $\Lambda_{1.4}$ from GW170817~\citep{abbo18}.}
	\label{k2lam}
\end{figure*}

The detection of the binary neutron star merger event GW170817 has provided observational constraints 
on the tidal deformation of neutron stars. The tidal deformability $\Lambda$ is determined by 
the EOS via both the stellar compactness $C=M/R$ and the tidal Love number $k_2$~\citep{hind08}，
as given by $\Lambda=2k_2 / \left(3C^5\right)$~\citep{ji19}. 
The resulting $k_2$ and $\Lambda$ are plotted against the neutron star mass $M$
in Figures~\ref{k2lam}(a) and~(b), respectively.
It is observed in Figure~\ref{k2lam}(a) that $k_2$ increases rapidly in the low-mass region, 
reaches its maximum at approximately $0.7$ to $0.9 \, M_{\odot}$, and then gradually decreases 
with increasing stellar mass $M$. The overall $k_2$ curves of TM1m*, TM1m, and SFHo 
lie below those of TM1 and NL3, which is attributed to their relatively soft EOSs. 
Generally, a soft EOS predicts a smaller stellar radius, resulting in 
a larger compactness $C$ and thus weakening the deformability under an external tidal field, 
which yields a smaller $k_2$. The softer EOSs of TM1m*, TM1m, and SFHo ultimately result in 
smaller tidal deformability $\Lambda$, as shown in Figure~\ref{k2lam}(b). 
In contrast, the stiffer EOSs of TM1 and NL3 yield larger $k_2$ and $\Lambda$, and 
their $\Lambda_{1.4}$ values exceed the observational constraints from GW170817~\citep{abbo18}.

\section{Summary} \label{sec:4}

In this work, we have investigated the influence of the effective nucleon mass and the corresponding
symmetry energy on the properties of pasta phases as well as the unified EOS of neutron stars.
The nonuniform pasta structure has been described within the self-consistent Thomas--Fermi approximation.
The calculations have been performed within the RMF approach employing six typical parameter sets:
TM1m*, TM1m, TM1e, TM1, NL3, and SFHo. 
Among these RMF models, the TM1m*, TM1m, and TM1e models are derived from the original TM1 model,
collectively referred to as the TM1-family models. 
The TM1m* and TM1m models have a larger effective mass of $M^{\ast}/M \sim 0.8$, 
while the TM1e and TM1 models have relatively smaller values of $M^{\ast}/M \sim 0.63$.
In terms of symmetry energy, the original TM1 model exhibits rather large symmetry 
energy $E_{\mathrm{sym}} = 36.9\,\mathrm{MeV}$ and slope parameter $L = 111\,\mathrm{MeV}$
at saturation density. In contrast, the corresponding values are 
$E_{\mathrm{sym}}=31.4 \, \mathrm{MeV}$ and $L = 40\,\mathrm{MeV}$ for the TM1m*, TM1m, and TM1e models.
In the TM1-family models, the isoscalar saturation properties, 
such as the saturation density $n_0$, energy per particle $E/A$, and incompressibility $K$,
remain the same, while the effective mass $M^{\ast}/M$ and the behaviors of symmetry
energy ($E_{\mathrm{sym}}$ and $L$) are significantly different.
By comparing the results of TM1m* and TM1m with those of TM1e, we can analyze the impact 
of the effective nucleon mass, whereas the comparison between TM1e and TM1 can reveal 
the influence of symmetry energy $E_{\mathrm{sym}}$ and its slope $L$.
For further examination of model dependence, we have also considered two other widely used models: 
the NL3 model with a large slope parameter of $L=118$ MeV, comparable to that of TM1, 
and the SFHo model known for its relatively large effective mass of $M^{\ast}/M \sim 0.76$,
comparable to those of TM1m* and TM1m.

Our results have indicated that the symmetry energy and its slope are crucial for determining 
the onset of pasta phases and their geometric configurations, whereas the effective nucleon mass 
has a relatively weak influence. At lower densities, the symmetry energy curves of TM1m* and TM1m 
lie below that of TM1e, leading to a lower proton fraction $Y_p$ and a smaller size of the 
Wigner--Seitz cell. Furthermore, nonspherical configurations emerge at a higher density 
for TM1m* and TM1m than for TM1e, owing to their lower symmetry energy at subsaturation densities.
A similar trend has also been observed in the SFHo model. 
On the other hand, the TM1 and NL3 models undergo a direct transition from
the droplet phase to uniform matter at relatively low densities, such that nonspherical phases 
cannot emerge prior to the crust–core transition. 
This phenomenon is attributed to the larger symmetry energy slope $L$ of the TM1 and NL3 models,
since the crust-core transition density is negatively correlated with $L$.

We have constructed unified EOSs by connecting the inner crust EOS obtained
within the TF approximation to the core EOS at the crust-core transition, 
in which the same RMF model has been adopted for both the inner crust and core EOSs.
With the unified EOSs obtained, we have explored the influence of the effective nucleon mass 
and the symmetry energy on the mass‑radius relation and the tidal deformability of neutron stars.
At high densities, the impact of the effective mass becomes more significant and 
a larger effective mass suppresses the kinetic energy contribution to the pressure, 
leading to a softer EOS. Consequently, the radius $R_{1.4}$ and the tidal deformability $\Lambda_{1.4}$
obtained in TM1m*, TM1m, and SFHo are smaller than those obtained in TM1e.
These results are compatible with current observational constraints.
In contrast, the TM1 and NL3 models yield much larger values of $R_{1.4}$ and $\Lambda_{1.4}$,
owing to their larger symmetry energy slope $L$, and these values exceed the observational 
constraints from the GW170817 event.
We emphasize that both the effective nucleon mass and the symmetry energy can affect
the properties of nuclear pasta phases in the inner crust as well as the bulk properties 
of neutron stars. Furthermore, these effects at finite temperature remain to be investigated.

\section{Acknowledgments}

This work was supported in part by the National Natural Science Foundation of China (grant No. 12475149) and the Guangdong Basic and Applied Basic Research Foundation (grant No. 2024A1515010911).

\bibliographystyle{aasjournalv7.1}
	
\bibliography{refs-LiSY}

\end{document}